\documentclass[%
  a4paper, twocolumn, pra, superscriptaddress, amsmath, amssymb, floatfix
]{revtex4-1}
\usepackage{graphicx}
\usepackage{epstopdf}
\usepackage[table]{xcolor}
\usepackage{braket}
\usepackage[colorlinks,bookmarksopen=false]{hyperref}
\usepackage[caption=false]{subfig}
\usepackage{bbm}
\usepackage{lipsum, blindtext}
\usepackage{siunitx}

\hypersetup{%
  pdfstartview={FitH},
  linkcolor=blue,
  citecolor=blue,
  filecolor=black,
  menucolor=black,
  urlcolor =black
}

\newcommand{\im}{\ensuremath{\textup{i}}}

\newcommand{\rp}{\ensuremath{\mathfrak{Re}}}

\newcommand{\op}[1]{\ensuremath{\mathsf{#1}}}
\newcommand{\lop}[1]{\ensuremath{\mathcal{#1}}}

\newcommand{\opnorm}[1]{\left\lVert#1\right\rVert}

\newcommand{\pulse}{\mathcal{E}}

\newcommand{\myvec}[1]{\boldsymbol{#1}}

\newcommand{\dd}{\ensuremath{\mathrm{d}}}

\begin{document}

\title{%
  Optimally controlled quantum discrimination and estimation
}

\author{Daniel Basilewitsch}
\affiliation{%
  Theoretische Physik, Universit\"{a}t Kassel, D-34132 Kassel, Germany
}

\author{Haidong Yuan}
\email{hdyuan@mae.cuhk.edu.hk}
\affiliation{%
  Department of Mechanical and Automation Engineering, The Chinese University of
  Hong Kong, Shatin, Hong Kong
}

\author{Christiane P. Koch}
\email{christiane.koch@fu-berlin.de}
\affiliation{%
  Theoretische Physik, Universit\"{a}t Kassel, D-34132 Kassel, Germany
}
\affiliation{%
  Dahlem Center for Complex Quantum Systems and Fachbereich Physik, Freie
  Universit\"{a}t Berlin, Arnimallee 14, D-14195 Berlin, Germany
}

\date{\today}

\begin{abstract}
  Quantum discrimination and estimation are pivotal for many quantum
  technologies, and their performance depends on the optimal choice of probe
  state and measurement. Here we show that their performance can be further
  improved by suitably tailoring the pulses that make up the interferometer.
  Developing an optimal control framework and applying it to the discrimination
  and estimation of a magnetic field in the presence of noise, we find an
  increase in the overall achievable state distinguishability. Moreover, the
  maximum distinguishability can be stabilized for times that are more than an
  order of magnitude longer than the decoherence time.
\end{abstract}

\maketitle

%===============================================================================
\section{Introduction}
Quantum control has become a very versatile tool for quantum
technologies~\cite{EPJD.69.279,KochJPCM16}, including quantum
computation~\cite{PRA.68.062308, TommasoPRA04, GoerzJPB11, GoerzNPJQI17} and
quantum simulation~\cite{CuiQST17, OmranSci20}. It is based on defining a figure
of merit which quantifies how well the desired target is reached and which is
taken to be a functional of yet unknown external fields~\cite{EPJD.69.279,
KochJPCM16}. Minimization, respectively maximization, of the functional yields
pulse shapes for the external fields that drive the system to a target state or
that implement a desired gate operation~\cite{EPJD.69.279, KochJPCM16}. Various
methods are now routinely being used to derive the pulse shapes, including both
gradient-based optimization methods such as GRadient Ascent Pulse Engineering
(GRAPE)~\cite{KhanejaJMR05}, Krotov's method~\cite{PRA.68.062308,
JCP.136.104103, GoerzSciPost19}, or the Gradient Optimization of Analytic
conTrols (GOAT) algorithm~\cite{MachnesPRL18}, as well as gradient-free
optimization such as the Chopped RAndom-Basis (CRAB) method~\cite{DoriaPRL11,
CanevaPRA11}.

The situation for applying quantum control, when compared with that in quantum
computation or quantum simulation, is a bit different in quantum discrimination
and quantum estimation, where the objective often does not involve
a well-defined target state or gate~\cite{HELS67, HOLE82, Yuen1975,
giovannetti2011advances, giovannetti2006quantum, braunstein1996generalized,
Fujiwara2008, escher2012general, demkowicz2014using, demkowicz2012elusive,
huelga1997improvement, chin2012quantum, Berry2015, Alipour2014, Beau2017,
Liu2019, tsang2016quantum}. For example, in quantum discrimination, the target
is to distinguish a discrete set of quantum states or channels~\cite{HELS67,
HOLE82, Ogawa2000, Ogawa2004, Hayashi2002, QChernoff2007, Audenaert2008,
Hiai1991, Hayashi2007, Acin2001, Mauro2001, Duan2007, Cheng2012, ChiribellaDP08,
DuanFY09, Harrow2010, Duan2008}. Instead of driving the system to a fixed
target, the control objective is to make the states more distinguishable to each
other, since -- intuitively -- the error gets smaller when the states become
more distinguishable. This is similar in quantum estimation~\cite{HELS67,
HOLE82, Yuen1975, giovannetti2011advances, giovannetti2006quantum,
braunstein1996generalized, Fujiwara2008, escher2012general, demkowicz2014using,
demkowicz2012elusive, huelga1997improvement, chin2012quantum, Berry2015,
Alipour2014, Beau2017, Liu2019, tsang2016quantum}, where the task is to estimate
the value of an unknown parameter encoded in the quantum dynamics. The error of
the estimation gets smaller when the states evolved with different values of the
parameter are more distinguishable.

Quantum discrimination and quantum estimation underlie many applications in
quantum information science, including quantum hypothesis testing, quantum
detection and quantum sensing. While quantum control has been employed to
improve the precision in quantum estimation~\cite{yuan2015optimal,
yuan2016sequential, Xu2019, Liu2017, Liu2017control, Pang2017, Hou19control,
naghiloo2017achieving, Predko2020, Mirkin2019}, the use of quantum control in
quantum discrimination remains scarce~\cite{Childs2000, Chen2019,
PhysRevA.98.043421}. This is so despite the fact that one may expect quantum
control to help identify fundamental performance bounds of quantum
discrimination, similar to those found for quantum
computation~\cite{CanevaPRL09, GoerzNPJQI17}, or derive pulse shapes for
improved performance with direct relevance to experiments~\cite{OmranSci20,
LarrouyPRX20}. All that is required is to adapt the quantum optimal control
toolbox to the specific use case of quantum discrimination.

Here, we develop a unified framework of optimal quantum control for quantum
discrimination and quantum estimation. We employ the distance between two states
that underwent different dynamics, more specifically that evolved under slightly
different magnetic field strengths, as the figure of the merit. In the limit of
the difference in field strength going to zero, optimizing this figure of merit
becomes equivalent to optimizing the quantum Fisher information. We use quantum
optimal control to maximize the distance between the two states by shaping the
external fields that make up the interferometer. Intuitively, this can be
understood as tailoring the external field to drive the states evolving under
different dynamics away from each other, instead of towards a common target.
Since both states depend on the control, the distance between them is typically
not a linear function, which is different from the case of a fixed target.
Krotov's method for quantum optimal control~\cite{JCP.136.104103,
GoerzSciPost19} can be used in such a case. We employ it here to optimize
discrimination and estimation of a magnetic field in the presence of noise,
increasing the performance compared to the standard scheme based on a Ramsey
interferometer. Our work thus contributes a quantum control perspective to
current efforts for improving quantum sensing protocols based on Ramsey
interferometry, using squeezed~\cite{Schulte2020} or
anticoherent~\cite{Martin2019} states, variable detuning of the
pulses~\cite{Sadzek2019}, or machine learning of the complete
protocol~\cite{PhysRevLett.124.060402}.

The article is organized as follows. We introduce the figure of merit for
discrimination and the estimation in Sec.~\ref{sec:model}, and then present the
quantum control method to optimize this figure of merit. In
Sec.~\ref{sec:results}, we apply the method to the discrimination and the
estimation of the magnetic fields to demonstrate the feature and advantages of
the control. We summarize our findings in Sec.~\ref{sec:concl}.

%===============================================================================
\section{Model and Control Problem}
\label{sec:model}
We consider the dynamics described by the Gorini-Kossakowski-Sudarshan-Lindblad
master equation~\cite{Breuer},
\begin{align} \label{eq:LvN}
  \frac{\dd}{\dd t} \op{\rho}_{m}(t)
  &=
  - \im \left[\op{H}_{m}(t), \op{\rho}_{m}(t)\right]
  \notag \\
  &\quad
  +
  \sum_{k}
  \gamma_{k} \Big(%
    \op{L}_{k} \op{\rho}_{m}(t) \op{L}_{k}^{\dagger}
    - \frac{1}{2} \big\{%
      \op{L}_{k}^{\dagger} \op{L}_{k}, \op{\rho}_{m}(t)
    \big\}
  \Big)
  \notag \\
  &=
  \lop{L}_{m}(t) \op{\rho}_{m}(t),
\end{align}
where
\begin{align}
 \op{H}_{m}(t)
  =
  \op{H}_{\mathrm{d},m} + \op{H}_{\mathrm{c}}(t),
\end{align}
is the Hamiltonian. $\op{H}_{\mathrm{d},m}$ describes the drift and
$\op{H}_{\mathrm{c}}(t)$ describes the coupling to an external drive,
$\op{L}_{k}$ are the Lindblad operators with the decay rates $\gamma_{k}$.

For quantum discrimination, we want to distinguish between two possible
Hamiltonians, $\op{H}_{\mathrm{d},1}$ and $\op{H}_{\mathrm{d},2}$, while for
quantum estimation, $\op{H}_{\mathrm{d},m}$ depends on a continuous parameter
which we want to estimate the value. In both cases $\op{H}_{\mathrm{d},m}$ can
not be measured directly, the discrimination (estimation) is achieved by the
measurement of the time evolved state $\op{\rho}_{m}(T)$ starting from an
initial state $\op{\rho}_{\mathrm{in}} = \ket{\Psi_{\mathrm{in}}}
\bra{\Psi_{\mathrm{in}}}$. For the discrimination of two Hamiltonians, the two
states $\op{\rho}_{1}(T)$ or $\op{\rho}_{2}(T)$ should be made as
distinguishable as possible. In contrast, for the estimation the precision can
also be connected to the distinguishability of the states that are evolved under
two neighboring Hamiltonian with $\op{H}_{\mathrm{d},1} = H(B-\delta B/2)$ and
$\op{H}_{\mathrm{d},2} = H(B+\delta B/2)$, where $\delta B$ is an infinitesimally
small shift~\cite{Brau94}. The difference between the discrimination and the
estimation is the figure of merit. The figure of merit for the discrimination is
typically taken as the success probability $P_{\mathrm{succ}}$ to distinguish
the two final states $\op{\rho}_{1}(T)$ and $\op{\rho}_{2}(T)$, which can be
related to the trace distance $D_{\mathrm{tr}}$ as~\cite{HELS67}
\begin{align} \label{eq:P_succ}
  P_{\mathrm{succ}}
  =
  \frac{1}{2}\bigg(%
    1 + D_{\mathrm{tr}} \left(\rho_1(T), \rho_2(T)\right)
    %\frac{1}{2} \left\lVert\rho_1(T)-\rho_2(T)\right\rVert_1
  \bigg)
\end{align}
where
\begin{align} \label{eq:tr_dist}
  D_{\mathrm{tr}}\left(\op{\rho}_{1}, \op{\rho}_{2}\right)
  =
  \frac{1}{2} \opnorm{\op{\rho}_{1} - \op{\rho}_{2}}
  _{\mathrm{tr}}
  \in [0,1],
\end{align}
$\opnorm{\op{\rho}}_{\mathrm{tr}} = \mathrm{Tr}\{\sqrt{\op{\rho}^{\dagger}
\op{\rho}}\}$. The figure of merit for the estimation is typically taken as the
precision, which can be calibrated by the quantum Cramer-Rao bound as
$E[(\hat{B}-B)^2] \geq \frac{1}{R \mathcal{F}_{\mathrm{Q}}}$, where
$E[(\hat{B}-B)^2]$ is the variance of an unbiased estimator $\hat{B}$, $R$ is
the number of repetition of the experiments and $\mathcal{F}_{\mathrm{Q}}$ is
the quantum Fisher information which determines the precision limit. Under the
two Hamiltonian $\op{H}_{\mathrm{d},1} = H(B-\delta B/2)$ and
$\op{H}_{\mathrm{d},2} = H(B+\delta B/2)$, the quantum Fisher information can be
related to the Bures distance $D_{\mathrm{bures}}$ between $\op{\rho}_{1}(T)$
and $\op{\rho}_{2}(T)$ as~\cite{Brau94}
\begin{align} \label{eq:qfi}
  \mathcal{F}_{\mathrm{Q}}
  =
  \frac{%
    4 D_{\mathrm{bures}}^{2}\left(\op{\rho}_{1}, \op{\rho}_{2}\right)
  }{%
    (\delta B)^{2}
  },
\end{align}
where the Bures distance between two states is defined
as~\cite{Jozsa.JModOpt.41.2315}
\begin{align} \label{eq:bures_dist}
  D_{\mathrm{bures}}^{2}\left(\op{\rho}_{1}, \op{\rho}_{2}\right)
  =
  2 - 2 \mathrm{Tr}\left\{\sqrt{%
      \sqrt{\op{\rho}_{1}} \op{\rho}_{2} \sqrt{\op{\rho}_{1}}
  }\right\}.
\end{align}

We consider distinguishing two Hamiltonians, $H_{\mathrm{d},1} = B_{1}
\op{\sigma}_{\mathrm{z}}/2$ and $H_{\mathrm{d},2} = B_{2}
\op{\sigma}_{\mathrm{z}}/2$. The discrimination of the two Hamiltonians can be
related to the estimation when $B_{1} = B - \delta B/2$ and $B_{2} = B + \delta
B/2$, which corresponds to the estimation of the strength of a magnetic field
oriented along the $z$-axis.

We first compare two protocols for the discrimination --- the standard Ramsey
protocol and the protocol employing optimized control fields. Each protocol
starts with preparing the qubit in the initial state $\op{\rho}_{\mathrm{in}}
= \ket{\Psi_{\mathrm{in}}} \bra{\Psi_{\mathrm{in}}}$ and is based on deducing
whether the field is $B_{1}$ or $B_{2}$ by means of measuring its time-evolved
state, $\op{\rho}_{m}(T)$. The Ramsey scheme is to prepare an initial state on
the Bloch sphere's equator and let it subsequently evolve under the constant
drift $\op{H}_{\mathrm{d},m}$, i.e., $\op{H}_{\mathrm{c}}(t)=0$. In contrast,
the optimized protocol will in addition employ time-dependent fields, i.e.,
$\op{H}_{\mathrm{c}}(t) \neq 0$. These control fields are optimized to make the
two states $\op{\rho}_{1}(T)$ and $\op{\rho}_{2}(T)$ as distinguishable as
possible. In other words, the optimized control fields need to maximize the
distance measure $D(\op{\rho}_{1}, \op{\rho}_{2})$. For the discrimination, the
distance is the trace distance~\eqref{eq:tr_dist}, since it is directly related
to the successful probability of the discrimination, cf. Eq.~\eqref{eq:P_succ}.
If expressed in terms of the Bloch vectors $\myvec{r}_{1}$ and $\myvec{r}_{2}$
for states $\op{\rho}_{1}$ and $\op{\rho}_{2}$, it reads $D_{\mathrm{tr}}
\left(\op{\rho}_{1}, \op{\rho}_{2}\right) = \lVert \myvec{r}_{1} - \myvec{r}_{2}
\rVert / 2$ with $\lVert \cdot \rVert$ the Euclidean vector
norm~\cite{NielsenChuang}.
% Note to self: I write here everything in terms of Bloch vectors expanded in
% the standard Pauli matrices. In the program, I use normalized Pauli matrices
% such that < S_i, S_i > = 1.
Thus, the trace distance coincides with the geometric distance between the Bloch
vectors $\myvec{r}_{1}$ and $\myvec{r}_{2}$ and maximal distinguishability is
achieved iff $\myvec{r}_{1}$ and $\myvec{r}_{2}$ are on opposite points on the
Bloch sphere. Hence, the maximization of $D_{\mathrm{tr}}$ will be our physical
goal for the discrimination.

The presence of the drive Hamiltonian $\op{H}_{\mathrm{c}}(t)$ allows to
influence the evolution of $D_{\mathrm{tr}}$. We make the general assumption
\begin{align}
  \op{H}_{\mathrm{c}}(t)
  =
  \frac{1}{2} \Big[
      \pulse_{\mathrm{x}}(t) \op{\sigma}_{\mathrm{x}}
    + \pulse_{\mathrm{y}}(t) \op{\sigma}_{\mathrm{y}}
    + \pulse_{\mathrm{z}}(t) \op{\sigma}_{\mathrm{z}}
  \Big],
\end{align}
where $\pulse_{\mathrm{x}}(t), \pulse_{\mathrm{y}}(t), \pulse_{\mathrm{z}}(t)
\in \mathbb{R}$ are control fields that couple via $\op{\sigma}_{\mathrm{x}}$,
$\op{\sigma}_{\mathrm{y}}$ and $\op{\sigma}_{\mathrm{z}}$ to the qubit,
respectively. Note that while $\op{H}_{\mathrm{c}}(t)$ is identical for both
Hamiltonians $\op{H}_{1}(t)$ and $\op{H}_{2}(t)$, it influences the dynamics
differently in the two cases due to the difference in the drift Hamiltonians.
It can thus be used to maximize $D_{\mathrm{tr}}$. The presence of
$\op{H}_{\mathrm{c}}(t)$ thus turns the discrimination problem into a control
problem, seeking to answer the question how to choose the three fields
$\pulse_{\mathrm{x}}(t)$, $\pulse_{\mathrm{y}}(t)$ and $\pulse_{\mathrm{z}}(t)$
such that $D_{\mathrm{tr}}$ is maximized at time $T$ when the state
$\op{\rho}_{m}(T)$ is measured.

\begin{figure*}[tb]
  \centering
  \includegraphics{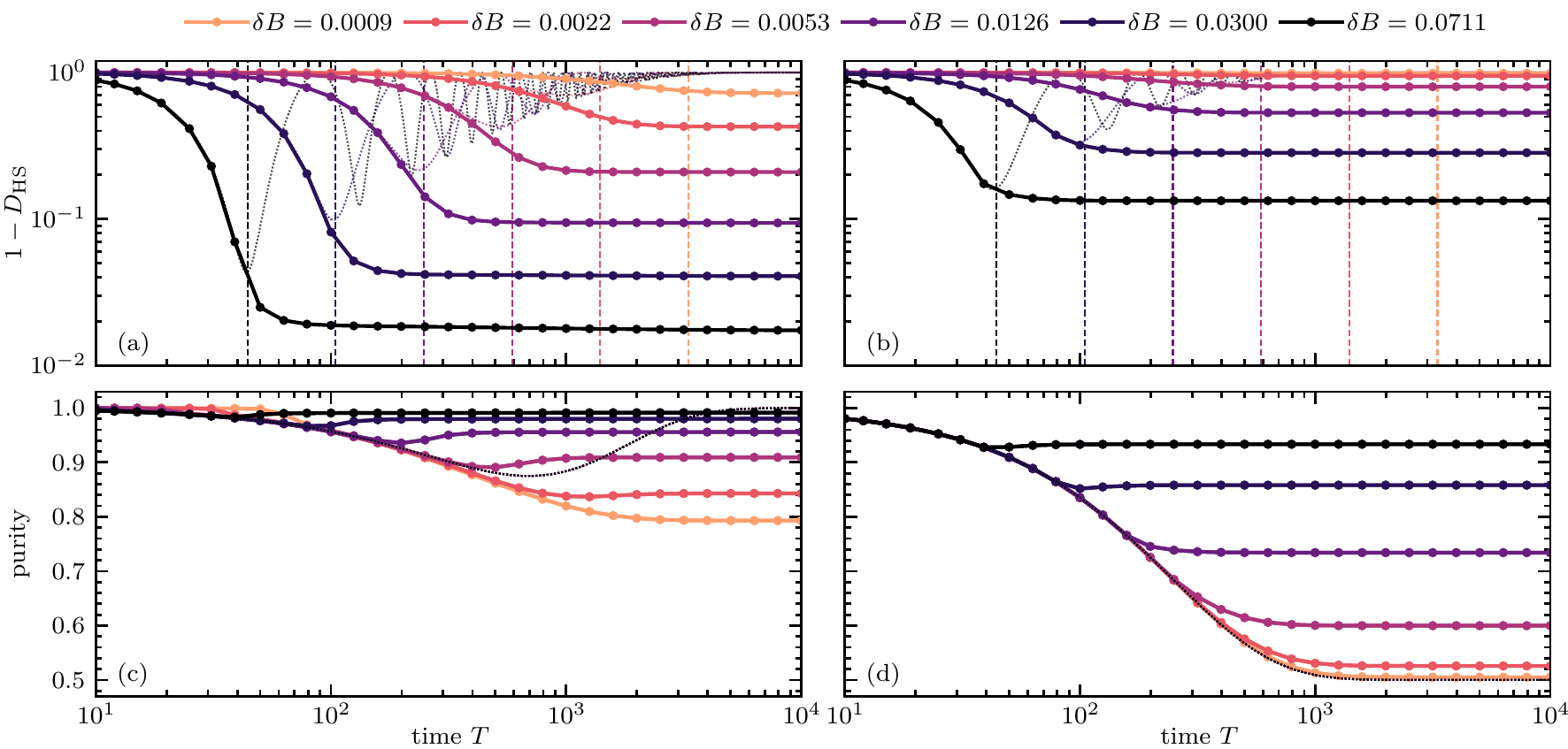}
  \caption{%
    Improvement of the state distinguishability under the optimized control
    fields. The upper graphs show the indistinguishability $1-D_{\mathrm{HS}}$
    as a function of protocol duration in case of (a) relaxation with
    $T_{1}=1000$ and (b) pure dephasing with $T_{2}=1000$. The dotted lines
    correspond to the Ramsey protocol whereas the markers indicates the
    reachable value of $1-D_{\mathrm{HS}}$ under the optimized control fields at
    the respective final time $T$. The vertical lines indicate the quantum speed
    limit given by Eq.~\eqref{eq:qsl}. Panels (c) and (d) show the purity of the
    two states corresponding to the dynamics of panels (a) and (b),
    respectively. Note that both states have almost identical purity, hence
    there is just one visible solid line per $\delta B$ that indicates the
    purity of the final states under the optimized control fields.
  }
  \label{fig:dist_all}
\end{figure*}

We derive suitable control fields employing optimal control
theory~\cite{EPJD.69.279}. To this end, we introduce the optimization functional
\begin{align} \label{eq:J}
  J\left[%
    \left\{\op{\rho}_{m}\right\}, \left\{\pulse_{k}\right\}
  \right]
  &=
  J_{T}\left[\left\{\op{\rho}_{m}(T)\right\}\right]
  \notag \\
  &\quad+ \int_{0}^{T} \text{d}t\,
    g\left[%
      \left\{\op{\rho}_{m}(t)\right\}, \left\{\pulse_{k}(t)\right\}, t
    \right],
\end{align}
where $J_{T}$ is the relevant figure of merit that quantifies the failure
probability or error at final time $T$ and $g$ captures additional running costs
at intermediate times. The sets $\{\op{\rho}_{m}\}$ and $\{\pulse_{k}\}$ are
forward propagated states and control fields, respectively, here given by
$\{\op{\rho}_{1}, \op{\rho}_{2}\}$ and $\{\pulse_{\mathrm{x}},
\pulse_{\mathrm{y}}, \pulse_{\mathrm{z}}\}$. Equation~\eqref{eq:J} describes the
most general form to represent an optimization functional and therefore
constitutes the standard ansatz to formulate an optimization
target~\cite{DAlessandro}. For the task of maximizing $D_{\mathrm{tr}}$, we
choose $J_{T}$ as
\begin{align} \label{eq:JT}
  J_{T}\left[\left\{\op{\rho}_{1}(T), \op{\rho}_{2}(T)\right\}\right]
  &=
  1 -
  D_{\mathrm{tr}}^{2} \left(\op{\rho}_{1}(T), \op{\rho}_{2}(T)\right)
  \notag \\
  &=
  1 -
  D_{\mathrm{HS}} \left(\op{\rho}_{1}(T), \op{\rho}_{2}(T)\right)
\end{align}
with $D_{\mathrm{HS}}$ the Hilbert-Schmidt
distance~\cite{Dodonov.JModOPt.47.633},
\begin{align}
  D_{\mathrm{HS}}\left(\op{\rho}_{1}, \op{\rho}_{2}\right)
  =
  \frac{1}{2} \left\langle
    \op{\rho}_{1} - \op{\rho}_{2},
    \op{\rho}_{1} - \op{\rho}_{2}
  \right\rangle
  \in [0,1],
\end{align}
where $\langle \op{A}, \op{B} \rangle = \mathrm{Tr}\{\op{A}^{\dagger} \op{B}\}$.
Note that the relation $D_{\mathrm{tr}}^{2} = D_{\mathrm{HS}}$ only holds for
qubits in which case maximization of $D_{\mathrm{tr}}$ and maximization of
$D_{\mathrm{HS}}$ are equivalent. Since both distances are appropriate measures
of state distinguishability, we choose $D_{\mathrm{HS}}$ for maximization in
optimal control, since it is more suitable for that
purpose~\cite{Xu.JChemPhys.120.6600, BasilewitschAQT} because it allows to build
analytical gradients with respect to the states $\op{\rho}_{1}$ and
$\op{\rho}_{2}$.

In the following we briefly describe our numerical algorithm of choice. We use
Krotov's method~\cite{AutomRemContr.60.1427}, an iterative and gradient-based
optimization technique, to minimize $J_{T}$, cf. Eq.~\eqref{eq:JT}. We achieve
the minimization of $J_{T}$ by minimizing the total functional $J$, cf.
Eq.~\eqref{eq:J}, assuming $g$ to take the form~\cite{PRA.68.062308}
\begin{align} \label{eq:g}
  g\left[%
    \left\{%
      \pulse_{\mathrm{x}}(t), \pulse_{\mathrm{y}}(t), \pulse_{\mathrm{z}}(t)
    \right\}
  \right]
  =
  \sum_{k=\mathrm{x},\mathrm{y},\mathrm{z}}
    \frac{\lambda_{k}}{S_{k}(t)}
    \left(\pulse_{k}(t) - \pulse_{k}^{\text{ref}}(t)\right)^{2},
\end{align}
where $\lambda_{k}$ is a numerical parameter, $S_{k}(t) \in (0,1]$ a shape
function and $\pulse_{k}^{\text{ref}}(t)$ a reference field.
Equation~\eqref{eq:g} is thereby a standard choice to control the pulse fluence
and should prevent the optimization to optimize towards unphysical pulse shapes.
With the choice of Eq.~\eqref{eq:g}, Krotov's method allows the derivation of
a closed form for the field update~\cite{JCP.136.104103},
\begin{widetext}
  \begin{align} \label{eq:pulse_update}
    \pulse_{k}^{(i+1)}(t)
    =
    \pulse_{k}^{\mathrm{ref}}(t)
    +
    \frac{S_{k}(t)}{\lambda_{k}} \rp\left\{%
      \sum_{m} \Braket{%
        \op{\chi}^{(i)}_{m}(t)\;,\;
        \frac{\partial \lop{L}\left[\left\{\pulse_{k'}\right\}\right]}{\partial
        \pulse_{k}} \Big|_{\{\pulse^{(i+1)}_{k'}(t)\}}
        \op{\rho}^{(i+1)}_{m}(t)
      }
    \right\},
  \end{align}
\end{widetext}
where the superscripts $(i)$ and $(i+1)$ indicate the previous and current
iteration, respectively. The states $\op{\rho}_{m}^{(i+1)}$ are determined by
solving
\begin{subequations} \label{eq:fw_states}
  \begin{align}
    \frac{\dd}{\dd t} \op{\rho}^{(i+1)}_{m}(t)
    &=
    \lop{L}^{(i+1)}(t) \op{\rho}^{(i+1)}_{m}(t),
    \\
    \op{\rho}^{(i+1)}_{m}(0)
    &=
    \op{\rho}_{\mathrm{in}}
  \end{align}
\end{subequations}
and the co-states $\op{\chi}_{m}^{(i)}$ by solving
\begin{subequations} \label{eq:co_states}
  \begin{align}
    \frac{\dd}{\dd t} \op{\chi}^{(i)}_{m}(t)
    &=
    \lop{L}^{\dagger (i)}(t) \op{\chi}^{(i)}_{m}(t),
    \\
    \op{\chi}^{(i)}_{m}(T)
    &=
    - \nabla_{\op{\rho}_{m}(T)} J_{T} \big|_{\{\op{\rho}^{(i)}_{m'}(T)\}}.
  \end{align}
\end{subequations}
The superscripts of the Liouvillians $\lop{L}$, cf. Eq.~\eqref{eq:LvN}, indicate
the respective iteration of the control fields. The reference field in
Eq.~\eqref{eq:pulse_update} is taken to be the field from the previous
iteration, i.e., $\pulse_{k}^{\mathrm{ref}}(t) = \pulse_{k}^{(i)}(t)$. Hence,
the running cost $g$ vanishes as the fields converge, and the total functional
$J$ essentially coincides with the relevant figure of merit $J_{T}$ that we seek
to minimize. See Ref.~\cite{JCP.136.104103} for a detailed description of
Krotov's method.

%===============================================================================
\section{Results and Discussion}
\label{sec:results}
The general time scale on which one can expect a given control task to be
feasible is an important property of the dynamics. For instance, for a control
problem where an initial state should be transferred into a given target state,
it is determined by the general speed of the evolution, typically set by the
Hamiltonian, and the distance between initial and target state. In our case,
however, we are interested in the relative distance $D_{\mathrm{HS}}$ between
the two time-evolved states $\op{\rho}_{1}(t)$ and $\op{\rho}_{2}(t)$ and not
into their distance with respect to the initial state $\op{\rho}_{\mathrm{in}}$.
Hence, the time scale on which $D_{\mathrm{HS}}$ increases is defined by their
relative speed of evolution. In detail, two different time scales are relevant
for the problem of maximizing $D_{\mathrm{HS}}$. On the one hand, there is
a quantum speed limit (QSL), i.e., a minimal time necessary to perfectly
distinguish the two states. Such a minimal time is defined for every physical
control task. Here, it is determined by $\delta B$ via the coherent part of the
dynamics and can be estimated by
\begin{align} \label{eq:qsl}
  T_{\mathrm{QSL}}
  =
  \frac{\pi}{\delta B}.
\end{align}
This is the minimal time required for perfect state distinguishability, i.e.,
$D_{\mathrm{HS}}=1$, in the Ramsey protocol and under the assumption of no
dissipation. On the other hand, dissipation continuously decreases
$D_{\mathrm{HS}}$, since it causes both states, $\op{\rho}_{1}(t)$ and
$\op{\rho}_{2}(t)$, evolving under $\op{H}_{1}(t)$ and $\op{H}_{2}(t)$ to evolve
towards the same steady state $\op{\rho}_{\mathrm{ss}}$. The time scale set by
the dissipation is, in contrast to the QSL, independent of $\delta B$. Since the
impact of relaxation and pure dephasing, characterized by $T_1$ and $T_2$,
respectively, is quite different, we consider them individually in the
following. This assumption is reasonable since in most physical settings, the
noise is either $T_{1}$ or $T_{2}$ dominated. We take $\ket{\Psi_{\mathrm{in}}}
= \ket{+} = (\ket{0} + \ket{1})/\sqrt{2}$ as initial state, in accordance with
the standard Ramsey scheme~\cite{HarocheRaimond}, i.e., in our dynamical
description, we do not account for the process preparing
$\ket{\Psi_{\mathrm{in}}}$.

Figure~\ref{fig:dist_all} shows the distinguishability $D_{\mathrm{HS}}$ as
a function of the protocol length $T$ for the Ramsey and optimized protocol. In
detail, the dotted lines in Fig.~\ref{fig:dist_all}(a) show the dynamics of
$1-D_{\mathrm{HS}}$ for the Ramsey protocol, i.e., $\op{H}_{\mathrm{c}}(t)=0$,
for several $\delta B$ under relaxation, i.e., a single Lindblad operator
$\op{L} = \ket{0}\bra{1}$ with $\gamma=1/T_{1}$. The dashed vertical lines
indicate the QSL of Eq.~\eqref{eq:qsl}. Starting at $D_{\mathrm{HS}}=0$ at
$T=0$, the distinguishability $D_{\mathrm{HS}}$ increases until it reaches the
maximum of $D_{\mathrm{HS}}^{\mathrm{max}}$ at approximately $T \approx
T_{\mathrm{QSL}}$. For times $T > T_{\mathrm{QSL}}$, the distinguishability
$D_{\mathrm{HS}}$ decreases exponentially as the relaxation causes
$\op{\rho}_{1}(t)$ and $\op{\rho}_{2}(t)$ to evolve towards the same
ground/steady state $\op{\rho}_{\mathrm{ss}} = \ket{0}\bra{0}$.

The decay of the state distinguishability due to relaxation can be completely
suppressed by using tailored, i.e., optimized, control fields. The markers in
Fig.~\ref{fig:dist_all}(a) show the reachable distinguishability
$D_{\mathrm{HS}}$ at the respective final time $T$ used in the optimization.
There are two interesting effects to notice. On the one hand, the reachable
maximum $D_{\mathrm{HS}}^{\mathrm{max}}$ increases compared to the Ramsey
protocol. Hence, in the presence of relaxation, optimized control fields allow
in general for better distinguishability despite a slightly longer protocol
duration (factor $\lesssim 2$) to reach $D_{\mathrm{HS}}^{\mathrm{max}}$. On the
other hand, the improvement in state distinguishability can be stabilized at
that maximally reachable distance against decay for protocol durations $T$ much
longer than the $T_{1}$ time. Figure~\ref{fig:dist_all}(a) demonstrates it for
times $T$ up to $10 \times T_{1}$ but suggests it should, in principle, be
feasible for even longer times.

\begin{figure}[tb]
  \centering
  \includegraphics{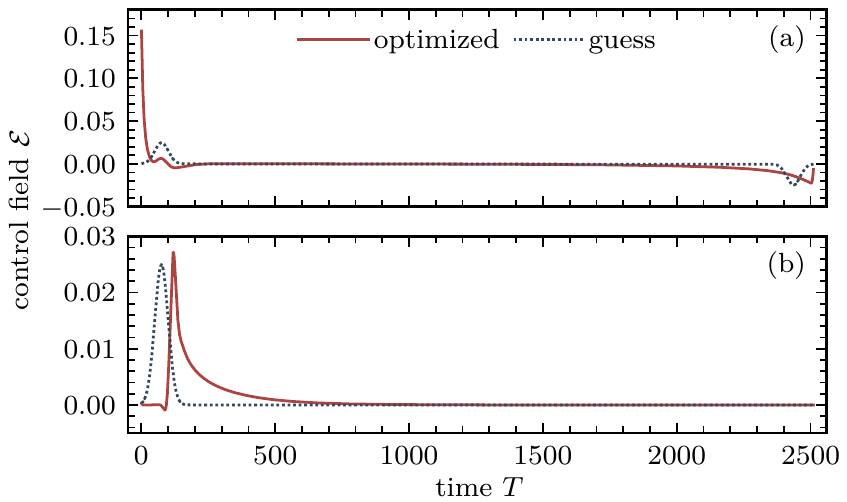}
  \caption{%
    Guess (dotted) and unconstrained optimized (solid) field for the case of (a)
    relaxation with the control field $\pulse_{\mathrm{y}}(t)$ and (b) pure
    dephasing where the control is $\pulse_{\mathrm{x}}(t)$. The Bloch sphere
    dynamics is depicted in Figs.~\ref{fig:single_bloch_trj}(a) and (b),
    respectively.
  }
  \label{fig:fields}
\end{figure}

\begin{figure}[tb]
  \centering
  \includegraphics{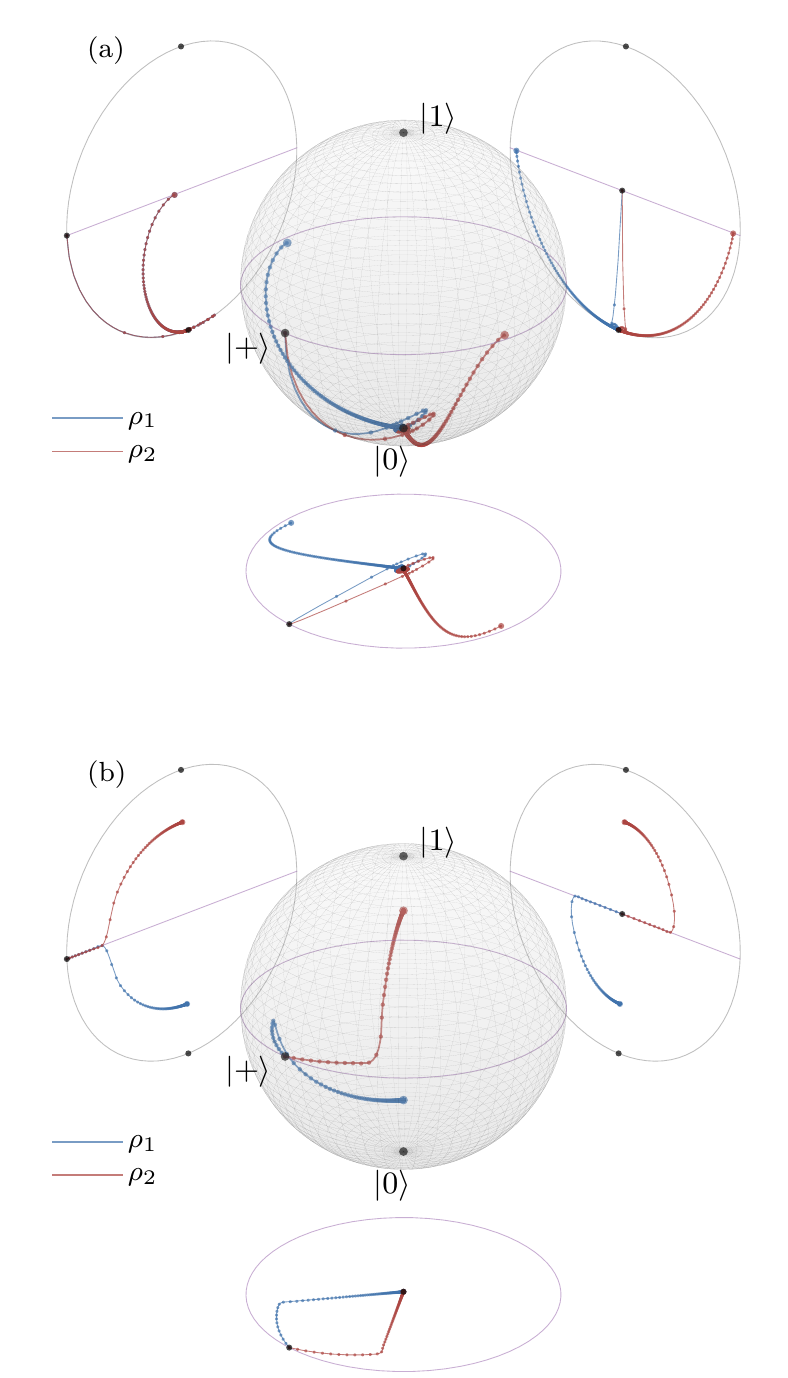}
  \caption{%
    Exemplary dynamics of the two states $\op{\rho}_{1}(t)$ and
    $\op{\rho}_{2}(t)$ under the optimized fields within the Bloch sphere for
    (a) relaxation and (b) pure dephasing. The parameters are $\delta
    B = 0.011$ with (a) $T_{1}=1000$ and (b) $T_{2}=1000$. The total time is
    $T=2511$ and the corresponding optimized fields are shown in
    Figs.~\ref{fig:fields}(a) and (b). The density of dots on each line
    indicates the speed of the evolution with a low density corresponding to
    high speed and vice versa.
  }
  \label{fig:single_bloch_trj}
\end{figure}

Figure~\ref{fig:dist_all}(c) shows the purities for states $\op{\rho}_{1}(t)$
and $\op{\rho}_{2}(t)$ corresponding to the data in Fig.~\ref{fig:dist_all}(a),
both for the Ramsey protocol (dotted lines) and at final time $T$ after an
evolution under the optimized control fields (markers). The dotted lines show
an intermediate purity loss in the Ramsey protocol due to the relaxation. The
final gain in purity for $t \rightarrow \infty$ is here a sign for the
incoherent process of both states approaching the same (pure) ground/steady
state. In contrast, the behavior of the purity in case of the improved and
stabilized $D_{\mathrm{HS}}$ depends on $\delta B$. While for larger $\delta B$
the loss of purity is avoided at all $T$ by the respective optimized control
fields, the improvement in case of small $\delta B$ comes along with a loss in
purity.

The improvement and stabilization of $D_{\mathrm{HS}}$ is achieved via a simple
control strategy which is most conveniently understood on the Bloch sphere, cf.
Fig.~\ref{fig:single_bloch_trj}(a). To this end, we choose the control field
$\pulse_{\mathrm{z}}$ such that it cancels the known $B$, i.e.,
$\pulse_{\mathrm{z}}(t) = - B$. This eliminates the fast, coherent oscillations
of $\myvec{r}_{1}(t)$ and $\myvec{r}_{2}(t)$ around the $z$-axis which do not
contribute to the distinguishability $D_{\mathrm{HS}}$. Furthermore, in order to
protect both states, $\myvec{r}_{1}(t)$ and $\myvec{r}_{2}(t)$, as much as
possible from the detrimental relaxation, i.e., prevent their vector norms from
shrinking, we kick both states from their initial position on the equator close
to the ground/steady state $\op{\rho}_{\mathrm{ss}} = \ket{0} \bra{0}$. This is
achieved by a $\pi/2$ like pulse via $\pulse_{\mathrm{y}}$ right at the
beginning of the protocol. The states will stay close to
$\op{\rho}_{\mathrm{ss}}$ for the largest part of the protocol where they evolve
effectively decoherence-free in the vicinity of $\op{\rho}_{\mathrm{ss}}$. For
the final measurement both states are transferred back to the equator by
a second, inverse $\pi/2$ like pulse.

Note that this strategy of protecting both states close to the ground/steady
state for as long as possible has been identified in steps. Initially, we
allowed the optimization of all three control fields $\pulse_{\mathrm{x}},
\pulse_{\mathrm{y}}, \pulse_{\mathrm{z}}$ and started optimizing without any
strategic choice for their guess fields. However, the above strategy (with only
slight deviations) has been identified even then. Its reduced version consists
of a constant $\pulse_{\mathrm{z}}$ and no $\pulse_{\mathrm{x}}$ at all such
that $\pulse_{\mathrm{y}}$ is the only time-dependent field that needs to be
optimized.

Figure~\ref{fig:fields}(a) shows, in an exemplary case, the guess and optimized
form of $\pulse_{\mathrm{y}}(t)$ when guiding the optimization with a guess
field that already incorporates the initial $\pi/2$ like kick in the beginning
and its inverse counterpart at the end~\footnote{It takes roughly 100 iterations
for Krotov's method to converge to these control fields. Employing e.g.\ the
QDYN library~\cite{QDYN}, this takes less than a minute on a standard desktop
computer.}. Compared to the guess field, the optimization increases the
intensity of the first kick such that the rotation from the initial equatorial
state $\op{\rho}_{\mathrm{in}} = \ket{+} \bra{+}$ towards
$\op{\rho}_{\mathrm{ss}}$ is carried out as fast as possible. The corresponding
dynamics on the Bloch sphere is shown in Fig.~\ref{fig:single_bloch_trj}(a).
After the first kick, the states remain most of the time close to the
ground/steady state $\op{\rho}_{\mathrm{ss}}$, which effectively protects them
from loosing purity.  The second, inverse kick is much smoother and transfers
the states symmetrically to the equatorial plane such that $D_{\mathrm{HS}}$
becomes maximal at $T$, i.e., the final time of measurement. The optimized field
in Fig.~\ref{fig:fields}(a) and its corresponding dynamics on the Bloch sphere,
cf.  Fig.~\ref{fig:single_bloch_trj}(a), have been picked as a representative of
an entire class of solutions for the problem of maximizing distinguishability in
the presence of relaxation. The exact details of the optimized control field and
corresponding dynamics differ depending on $\delta B$ and $T$, but the general
control strategy remains similar.

We now turn to the case of pure dephasing with Lindblad operator $\op{L}
= \op{\sigma}_{\mathrm{z}}$ and rate $\gamma=1/T_{2}$.
Figure~\ref{fig:dist_all}(b) shows the dynamics for the Ramsey protocol as
dotted lines. In comparison to the case of $T_{1}$ decay, cf.
Fig.~\ref{fig:dist_all}(a), pure dephasing has a more severe influence on
$D_{\mathrm{HS}}$ even if the decay rates are identical, $\gamma = 1/T_{2}
= 1/T_{1}$. But also in this case, optimization is capable of improving
$D_{\mathrm{HS}}$ over the Ramsey protocol --- again at the expense of longer
protocol durations (factor $\lesssim 2$). The effect of stabilizing
$D_{\mathrm{HS}}$ at the maximal reachable distance for times much longer than
the decay time can be observed as well. Nevertheless, the dynamics both in the
Ramsey protocol as well as under the optimized control fields look quite
different compared to relaxation. With pure dephasing no unique, single steady
state exists but rather a set of states, namely the coherence-free states given
by $\{\op{\rho}_{\mathrm{ss}} = p \ket{0}\bra{0} + (1-p) \ket{1}\bra{1} | p \in
[0,1]\}$, i.e., all states on the $z$-axis of the Bloch sphere. Since neither
the drift $\op{H}_{\mathrm{d},m}$ nor the dephasing cause a change of any
state's $z$-projection, the two states $\myvec{r}_{1}(t)$ and
$\myvec{r}_{2}(t)$, starting initially in the equatorial plane, precess around
the $z$-axis while loosing purity, i.e., shrink within the equatorial plane.
Hence, they evolve towards the Bloch sphere's center, i.e., the completely mixed
state. This is evidenced by the dotted lines in Fig.~\ref{fig:dist_all}(d),
which show the purity evolving towards $1/2$ under the Ramsey protocol.

An optimization of all three available control fields $\pulse_{\mathrm{x}}$,
$\pulse_{\mathrm{y}}$, $\pulse_{\mathrm{z}}$ again yields a simple control
strategy. Like in the case of relaxation, it can also be realized by a single
time-dependent control field, which is what for simplicity we discuss here. This
time, the time-dependent control is $\pulse_{\mathrm{x}}(t)$, while
$\pulse_{\mathrm{z}}(t) = - B$ again cancels the known field $B$ and
$\pulse_{\mathrm{y}}$ is not needed at all. Figure~\ref{fig:fields}(b) shows the
guess field for $\pulse_{\mathrm{x}}(t)$, which exhibits a peak at the
beginning. This peak is modified by the optimization such that it splits the two
states $\myvec{r}_{1}(t)$ and $\myvec{r}_{2}(t)$ within the equatorial plane as
a first step and then rotates them onto the $z$-axis in a second step, see
Fig.~\ref{fig:single_bloch_trj}(b) for the corresponding dynamics. Once the
states reach the $z$-axis, $\pulse_{\mathrm{x}}(t) \approx 0$ is essentially
turned off and the states become invariants of the dynamics which implies that
their distinguishability $D_{\mathrm{HS}}$ can essentially be preserved forever.
This readily explains the stabilization observed in Fig.~\ref{fig:dist_all}(b).
The respective optimized field and dynamics in Figs.~\ref{fig:fields}(b)
and~\ref{fig:single_bloch_trj}(b) again represent an example for the entire
class of solutions for the problem of maximal distinguishability in the case of
pure dephasing. The exact details depend again on $\delta B$ and $T$.

\begin{figure}[tb]
  \centering
  \includegraphics{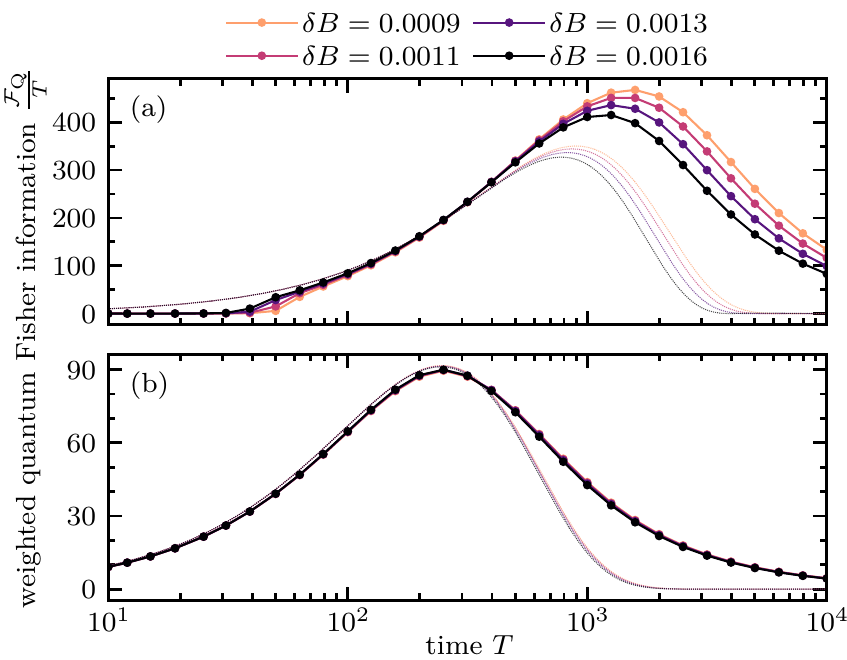}
  \caption{%
    Quantum Fisher information $\mathcal{F}_{\mathrm{Q}}$ (weighted by the
    protocol duration $T$) for small values of $\delta B$. Panels (a)
    corresponds to the case of relaxation presented in
    Fig.~\ref{fig:dist_all}(a) while panel (b) corresponds to the case of pure
    dephasing in Fig.~\ref{fig:dist_all}(b). The dotted lines indicate the
    values for the Ramsey protocol whereas the markers show the optimized
    results.
  }
  \label{fig:qfi}
\end{figure}

\begin{figure*}[tb]
  \centering
  \includegraphics{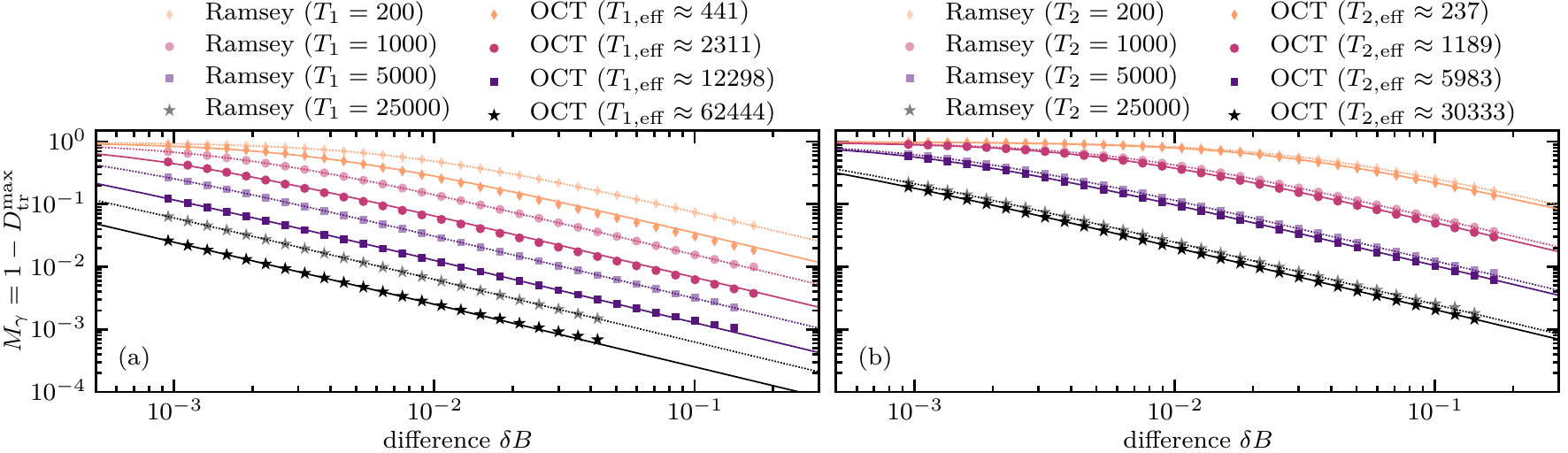}
  \caption{%
    Improvement of the state distinguishability in terms of effective decay
    rates. The plot shows the maximal trace distance
    $D_{\mathrm{tr}}^{\mathrm{max}}$, i.e., minimal $M_{\gamma}$, as a function
    of $\delta B$ for several (a) relaxation and (b) dephasing times. The values
    for the Ramsey scheme (opaque markers) follow the analytical prediction of
    Eq.~\eqref{eq:M}, shown as dotted lines. The non-opaque markers correspond
    to the optimized values of $D_{\mathrm{tr}}^{\mathrm{max}}$, i.e., numerical
    evaluation of Eq.~\eqref{eq:Mmin}. The solid lines are fits of the optimized
    values to Eq.~\eqref{eq:M} with fitting parameter (a)
    $\gamma=1/T_{1,\mathrm{eff}}$ and (b) $\gamma=4/T_{2,\mathrm{eff}}$.
  }
  \label{fig:dist_oct_improvement}
\end{figure*}

Next, we relate the improved distinguishability $D_{\mathrm{HS}}$ observed in
Fig.~\ref{fig:dist_all} to the quantum Fisher information
$\mathcal{F}_{\mathrm{Q}}$, cf. Eq.~\eqref{eq:qfi}. However, it depends on the
Bures distance $D_{\mathrm{bures}}$, which is a distance metric on the set of
density matrices, just as the trace distance $D_{\mathrm{tr}}$ or the
Hilbert-Schmidt distance $D_{\mathrm{HS}}$. Unlike the trace distance discussed
above, $D_{\mathrm{bures}}$ cannot be related to $D_{\mathrm{HS}}$, not even in
the case of qubits. Nevertheless, the increase of $D_{\mathrm{HS}}$ is expected
to increase $D_{\mathrm{bures}}$ as well~\cite{BasilewitschAQT}. For the
maximization of $D_{\mathrm{HS}}$, shown in Fig.~\ref{fig:dist_all}, this is in
fact true and $D_{\mathrm{bures}}$ is readily improved alongside
$D_{\mathrm{HS}}$.

Note that Eq.~\eqref{eq:qfi} is only valid for small $\delta B$. Moreover, it
needs to be weighted by the protocol duration $T$ in order to quantify the
amount of information that can be obtained per unit time for any given protocol.
Accordingly, Fig.~\ref{fig:qfi} shows the quantum Fisher information
$\mathcal{F}_{\mathrm{Q}}$ weighted by the protocol duration for small values of
$\delta B$. In the case of pure dephasing, cf. Fig.~\ref{fig:qfi}(b), there is
a small improvement in $D_{\mathrm{HS}}$, respectively $D_{\mathrm{bures}}$, for
the optimized protocol compared to the Ramsey protocol. This is, however, almost
completely canceled by the slightly longer protocol duration $T$. As a result,
the maximally reachable value of $\mathcal{F}_{\mathrm{Q}}/T$ is almost
identical for the Ramsey and optimized protocols. In contrast, for relaxation,
cf. Fig.~\ref{fig:qfi}(a), the significant improvement of $D_{\mathrm{HS}}$,
respectively $D_{\mathrm{bures}}$, realized by the optimized protocol gives rise
to an improvement of $\mathcal{F}_{\mathrm{Q}}/T$ despite the slightly longer
protocol duration $T$. We thus expect a metrological gain of the optimized
protocol compared to the Ramsey protocol.

So far, we only considered decay rates determined by $T_{1}=1000$ and
$T_{2}=1000$. However, since the dissipation sets a time scale for the control
task that is independent on the QSL set by $\delta B$, cf. Eq.~\eqref{eq:qsl},
it is natural to ask whether the control strategy that has been identified above
depends on the decay rates. To this end, we examine how the improvement of
$D_{\mathrm{HS}}$, respectively $D_{\mathrm{tr}} = \sqrt{D_{\mathrm{HS}}}$,
observed in Fig.~\ref{fig:dist_all} behaves for different relaxation and
dephasing times. In detail, we are interested in the behavior of
\begin{align} \label{eq:Mmin}
  M_{\gamma}(\delta B)
  &\equiv
  \underset{t}{\mathrm{min}} \left\{%
    1 - D_{\mathrm{tr}} \left(\op{\rho}_{1}(t), \op{\rho}_{2}(t)\right)
  \right\},
\end{align}
as a function of $\delta B$ and for various decay rates $1/T_{1}$ and $1/T_{2}$.
The function $M_{\gamma}$ measures, for a given $\delta B$, the maximally
reachable distinguishability $D_{\mathrm{tr}}^{\mathrm{max}}$, independent of
the time it takes to reach it. In other words, $M_{\gamma}(\delta B)
= 1 - D_{\mathrm{tr}}^{\mathrm{max}}$. If, for a given physical process, the
protocol duration is not crucial and only the maximally achievable state
distinguishability is of importance, $M_{\gamma}(\delta B)$ is the relevant
figure of merit. For the Ramsey protocol, Eq.~\eqref{eq:Mmin} can be solved
analytically to yield
\begin{align} \label{eq:M}
  M_{\gamma}(\delta B)
  &=
  1 - \Bigg[ \frac{(\delta B)^{2}}{(\delta B)^{2} + \gamma^{2}}
  \notag \\
  &\qquad \quad
  \times
  \exp\left\{%
    - \frac{\gamma}{\delta B}
    \arccos\left(
      \frac{\gamma^{2} - (\delta B)^{2}}{\gamma^{2} + (\delta B)^{2}}
    \right)
  \right\} \Bigg]^{1/2}
\end{align}
for relaxation with $\gamma=1/T_{1}$. For pure dephasing, the solution takes the
same form but differs by a factor of four, i.e., $\gamma=4/T_{2}$. The dotted
lines in Figs.~\ref{fig:dist_oct_improvement}(a) and (b) show $M_{\gamma}$ for
the Ramsey protocol for relaxation and pure dephasing, respectively. The
dotted lines perfectly fit the numerical values given by the opaque markers, as
expected for an analytical solution. For the dynamics under the optimized
control fields, we can evaluate Eq.~\eqref{eq:Mmin} numerically, cf.\ the
non-opaque markers in Fig.~\ref{fig:dist_oct_improvement}. Remarkably, these
show an almost identical functional dependence compared to the Ramsey scheme. We
therefore fit the data obtained for the optimized protocol to Eq.~\eqref{eq:M}
using effective relaxation and dephasing times as fitting parameters. This
yields the solid lines in Fig.~\ref{fig:dist_oct_improvement}, which indeed show
that $M_{\gamma}(\delta B)$ accurately describes the dependence also for the
optimized data points with effective decay times $T_{1,\mathrm{eff}}$ or
$T_{2,\mathrm{eff}}$, see the legends in Fig.~\ref{fig:dist_oct_improvement}.
This is in fact not obvious as the coherent dynamics of the Ramsey and optimized
protocol differ drastically, which makes the resemblance in their functional
behavior of $M_{\gamma}$ remarkable. For relaxation, the effective decay times
satisfy $T_{1,\mathrm{eff}}/T_{1} \approx 2.4$, whereas for pure dephasing, the
ratio is $T_{2,\mathrm{eff}}/T_{2} \approx 1.2$. Thus, the maximally reachable
distinguishability $D_{\mathrm{tr}}^{\mathrm{max}}$ behaves as though it would
have been measured by a Ramsey protocol with $2.4$ times longer $T_{1}$,
respectively $1.2$ times longer $T_{2}$ time, which greatly improves the
distinguishability. Given the protection strategy of the dynamics, the
prolongation of the decay times is not surprising, since the overall impact of
the dissipation onto the states is reduced.

%===============================================================================
\section{Conclusions}
\label{sec:concl}
In summary, we have studied how optimized control fields can help to improve the
distinguishability of two states of a qubit --- both of which evolve under
different drift but identical drive Hamiltonians while being exposed to either
relaxation or pure dephasing. Our results show two improvements with respect to
a standard Ramsey protocol for state discrimination.

First, optimized control fields increase the overall achievable state
distinguishability, at the expense of slightly longer protocol durations. When
comparing this improved state distinguishability against the prolonged protocol
duration, in the case of relaxation, we observe a metrological gain, evidenced
by the quantum Fisher information weighted by the protocol duration. In
contrast, both effects --- the improved state distinguishability and the
prolonged protocol duration --- roughly cancel in the case of pure dephasing.

Second, by utilizing optimize control fields, we are not only able to improve
the state distinguishability but also to stabilize it at its maximum for times
that are at least one order of magnitude longer than the decay times due to the
environmental noise. The control strategy utilizes decoherence-free subspaces in
all cases, where the states can be effectively stored and protected before being
separated right before their measurement. We find the required control fields to
be both simple and experimentally feasible.

Our study demonstrates the capabilities of optimal control to effectively reduce
the environments detrimental influence. For the considered state discrimination
problem and if compared to the standard Ramsey scheme, it reveals an alternative
protocol with improved noise resistance. Our results thus suggest to explore
state discrimination and its impact on quantum metrological applications from
a new perspective.

\begin{acknowledgments}
  We would like to thank Daniel M. Reich for fruitful discussions and gratefully
  acknowledge financial support from the Volkswagenstiftung (Grant No. 91004),
  DAAD (Grant No. 57513913 ), the Research Grants Council of Hong Kong (Grant
  No. 14308019) and the Research Strategic Funding Scheme of The Chinese
  University of Hong Kong (Grant No.  3133234).
\end{acknowledgments}

%\bibliography{refs}

\begin{thebibliography}{77}%
\makeatletter
\providecommand \@ifxundefined [1]{%
 \@ifx{#1\undefined}
}%
\providecommand \@ifnum [1]{%
 \ifnum #1\expandafter \@firstoftwo
 \else \expandafter \@secondoftwo
 \fi
}%
\providecommand \@ifx [1]{%
 \ifx #1\expandafter \@firstoftwo
 \else \expandafter \@secondoftwo
 \fi
}%
\providecommand \natexlab [1]{#1}%
\providecommand \enquote  [1]{``#1''}%
\providecommand \bibnamefont  [1]{#1}%
\providecommand \bibfnamefont [1]{#1}%
\providecommand \citenamefont [1]{#1}%
\providecommand \href@noop [0]{\@secondoftwo}%
\providecommand \href [0]{\begingroup \@sanitize@url \@href}%
\providecommand \@href[1]{\@@startlink{#1}\@@href}%
\providecommand \@@href[1]{\endgroup#1\@@endlink}%
\providecommand \@sanitize@url [0]{\catcode `\\12\catcode `\$12\catcode
  `\&12\catcode `\#12\catcode `\^12\catcode `\_12\catcode `\%12\relax}%
\providecommand \@@startlink[1]{}%
\providecommand \@@endlink[0]{}%
\providecommand \url  [0]{\begingroup\@sanitize@url \@url }%
\providecommand \@url [1]{\endgroup\@href {#1}{\urlprefix }}%
\providecommand \urlprefix  [0]{URL }%
\providecommand \Eprint [0]{\href }%
\providecommand \doibase [0]{http://dx.doi.org/}%
\providecommand \selectlanguage [0]{\@gobble}%
\providecommand \bibinfo  [0]{\@secondoftwo}%
\providecommand \bibfield  [0]{\@secondoftwo}%
\providecommand \translation [1]{[#1]}%
\providecommand \BibitemOpen [0]{}%
\providecommand \bibitemStop [0]{}%
\providecommand \bibitemNoStop [0]{.\EOS\space}%
\providecommand \EOS [0]{\spacefactor3000\relax}%
\providecommand \BibitemShut  [1]{\csname bibitem#1\endcsname}%
\let\auto@bib@innerbib\@empty
%</preamble>
\bibitem [{\citenamefont {Glaser}\ \emph {et~al.}(2015)\citenamefont {Glaser},
  \citenamefont {Boscain}, \citenamefont {Calarco}, \citenamefont {Koch},
  \citenamefont {K\"ockenberger}, \citenamefont {Kosloff}, \citenamefont
  {Kuprov}, \citenamefont {Luy}, \citenamefont {Schirmer}, \citenamefont
  {Schulte-Herbr\"uggen}, \citenamefont {Sugny},\ and\ \citenamefont
  {Wilhelm}}]{EPJD.69.279}%
  \BibitemOpen
  \bibfield  {author} {\bibinfo {author} {\bibfnamefont {S.~J.}\ \bibnamefont
  {Glaser}}, \bibinfo {author} {\bibfnamefont {U.}~\bibnamefont {Boscain}},
  \bibinfo {author} {\bibfnamefont {T.}~\bibnamefont {Calarco}}, \bibinfo
  {author} {\bibfnamefont {C.~P.}\ \bibnamefont {Koch}}, \bibinfo {author}
  {\bibfnamefont {W.}~\bibnamefont {K\"ockenberger}}, \bibinfo {author}
  {\bibfnamefont {R.}~\bibnamefont {Kosloff}}, \bibinfo {author} {\bibfnamefont
  {I.}~\bibnamefont {Kuprov}}, \bibinfo {author} {\bibfnamefont
  {B.}~\bibnamefont {Luy}}, \bibinfo {author} {\bibfnamefont {S.}~\bibnamefont
  {Schirmer}}, \bibinfo {author} {\bibfnamefont {T.}~\bibnamefont
  {Schulte-Herbr\"uggen}}, \bibinfo {author} {\bibfnamefont {D.}~\bibnamefont
  {Sugny}}, \ and\ \bibinfo {author} {\bibfnamefont {F.~K.}\ \bibnamefont
  {Wilhelm}},\ }\href {\doibase 10.1140/epjd/e2015-60464-1} {\bibfield
  {journal} {\bibinfo  {journal} {Eur. Phys. J. D}\ }\textbf {\bibinfo {volume}
  {69}},\ \bibinfo {pages} {279} (\bibinfo {year} {2015})}\BibitemShut
  {NoStop}%
\bibitem [{\citenamefont {Koch}(2016)}]{KochJPCM16}%
  \BibitemOpen
  \bibfield  {author} {\bibinfo {author} {\bibfnamefont {C.~P.}\ \bibnamefont
  {Koch}},\ }\href@noop {} {\bibfield  {journal} {\bibinfo  {journal} {J.
  Phys.: Condens. Matter}\ }\textbf {\bibinfo {volume} {28}},\ \bibinfo {pages}
  {213001} (\bibinfo {year} {2016})}\BibitemShut {NoStop}%
\bibitem [{\citenamefont {Palao}\ and\ \citenamefont
  {Kosloff}(2003)}]{PRA.68.062308}%
  \BibitemOpen
  \bibfield  {author} {\bibinfo {author} {\bibfnamefont {J.~P.}\ \bibnamefont
  {Palao}}\ and\ \bibinfo {author} {\bibfnamefont {R.}~\bibnamefont
  {Kosloff}},\ }\href {\doibase 10.1103/PhysRevA.68.062308} {\bibfield
  {journal} {\bibinfo  {journal} {Phys. Rev. A}\ }\textbf {\bibinfo {volume}
  {68}},\ \bibinfo {pages} {062308} (\bibinfo {year} {2003})}\BibitemShut
  {NoStop}%
\bibitem [{\citenamefont {Calarco}\ \emph {et~al.}(2004)\citenamefont
  {Calarco}, \citenamefont {Dorner}, \citenamefont {Julienne}, \citenamefont
  {Williams},\ and\ \citenamefont {Zoller}}]{TommasoPRA04}%
  \BibitemOpen
  \bibfield  {author} {\bibinfo {author} {\bibfnamefont {T.}~\bibnamefont
  {Calarco}}, \bibinfo {author} {\bibfnamefont {U.}~\bibnamefont {Dorner}},
  \bibinfo {author} {\bibfnamefont {P.~S.}\ \bibnamefont {Julienne}}, \bibinfo
  {author} {\bibfnamefont {C.~J.}\ \bibnamefont {Williams}}, \ and\ \bibinfo
  {author} {\bibfnamefont {P.}~\bibnamefont {Zoller}},\ }\href
  {http://link.aps.org/abstract/PRA/v70/e012306} {\bibfield  {journal}
  {\bibinfo  {journal} {Phys. Rev. A}\ }\textbf {\bibinfo {volume} {70}},\
  \bibinfo {pages} {012306} (\bibinfo {year} {2004})}\BibitemShut {NoStop}%
\bibitem [{\citenamefont {Goerz}\ \emph {et~al.}(2011)\citenamefont {Goerz},
  \citenamefont {Calarco},\ and\ \citenamefont {Koch}}]{GoerzJPB11}%
  \BibitemOpen
  \bibfield  {author} {\bibinfo {author} {\bibfnamefont {M.~H.}\ \bibnamefont
  {Goerz}}, \bibinfo {author} {\bibfnamefont {T.}~\bibnamefont {Calarco}}, \
  and\ \bibinfo {author} {\bibfnamefont {C.~P.}\ \bibnamefont {Koch}},\ }\href
  {\doibase 10.1088/0953-4075/44/15/154011} {\bibfield  {journal} {\bibinfo
  {journal} {J. Phys. B}\ }\textbf {\bibinfo {volume} {44}},\ \bibinfo {pages}
  {154011} (\bibinfo {year} {2011})}\BibitemShut {NoStop}%
\bibitem [{\citenamefont {Goerz}\ \emph {et~al.}(2017)\citenamefont {Goerz},
  \citenamefont {Motzoi}, \citenamefont {Whaley},\ and\ \citenamefont
  {Koch}}]{GoerzNPJQI17}%
  \BibitemOpen
  \bibfield  {author} {\bibinfo {author} {\bibfnamefont {M.~H.}\ \bibnamefont
  {Goerz}}, \bibinfo {author} {\bibfnamefont {F.}~\bibnamefont {Motzoi}},
  \bibinfo {author} {\bibfnamefont {K.~B.}\ \bibnamefont {Whaley}}, \ and\
  \bibinfo {author} {\bibfnamefont {C.~P.}\ \bibnamefont {Koch}},\ }\href
  {\doibase 10.1038/s41534-017-0036-0} {\bibfield  {journal} {\bibinfo
  {journal} {npj Quantum Inf.}\ }\textbf {\bibinfo {volume} {3}},\ \bibinfo
  {pages} {37} (\bibinfo {year} {2017})}\BibitemShut {NoStop}%
\bibitem [{\citenamefont {Cui}\ \emph {et~al.}(2017)\citenamefont {Cui},
  \citenamefont {van Bijnen}, \citenamefont {Pohl}, \citenamefont
  {Montangero},\ and\ \citenamefont {Calarco}}]{CuiQST17}%
  \BibitemOpen
  \bibfield  {author} {\bibinfo {author} {\bibfnamefont {J.}~\bibnamefont
  {Cui}}, \bibinfo {author} {\bibfnamefont {R.}~\bibnamefont {van Bijnen}},
  \bibinfo {author} {\bibfnamefont {T.}~\bibnamefont {Pohl}}, \bibinfo {author}
  {\bibfnamefont {S.}~\bibnamefont {Montangero}}, \ and\ \bibinfo {author}
  {\bibfnamefont {T.}~\bibnamefont {Calarco}},\ }\href {\doibase
  10.1088/2058-9565/aa7daf} {\bibfield  {journal} {\bibinfo  {journal} {Quantum
  Sci. Technol.}\ }\textbf {\bibinfo {volume} {2}},\ \bibinfo {pages} {035006}
  (\bibinfo {year} {2017})}\BibitemShut {NoStop}%
\bibitem [{\citenamefont {Omran}\ \emph {et~al.}(2019)\citenamefont {Omran},
  \citenamefont {Levine}, \citenamefont {Keesling}, \citenamefont {Semeghini},
  \citenamefont {Wang}, \citenamefont {Ebadi}, \citenamefont {Bernien},
  \citenamefont {Zibrov}, \citenamefont {Pichler}, \citenamefont {Choi},
  \citenamefont {Cui}, \citenamefont {Rossignolo}, \citenamefont {Rembold},
  \citenamefont {Montangero}, \citenamefont {Calarco}, \citenamefont {Endres},
  \citenamefont {Greiner}, \citenamefont {Vuleti{\'c}},\ and\ \citenamefont
  {Lukin}}]{OmranSci20}%
  \BibitemOpen
  \bibfield  {author} {\bibinfo {author} {\bibfnamefont {A.}~\bibnamefont
  {Omran}}, \bibinfo {author} {\bibfnamefont {H.}~\bibnamefont {Levine}},
  \bibinfo {author} {\bibfnamefont {A.}~\bibnamefont {Keesling}}, \bibinfo
  {author} {\bibfnamefont {G.}~\bibnamefont {Semeghini}}, \bibinfo {author}
  {\bibfnamefont {T.~T.}\ \bibnamefont {Wang}}, \bibinfo {author}
  {\bibfnamefont {S.}~\bibnamefont {Ebadi}}, \bibinfo {author} {\bibfnamefont
  {H.}~\bibnamefont {Bernien}}, \bibinfo {author} {\bibfnamefont {A.~S.}\
  \bibnamefont {Zibrov}}, \bibinfo {author} {\bibfnamefont {H.}~\bibnamefont
  {Pichler}}, \bibinfo {author} {\bibfnamefont {S.}~\bibnamefont {Choi}},
  \bibinfo {author} {\bibfnamefont {J.}~\bibnamefont {Cui}}, \bibinfo {author}
  {\bibfnamefont {M.}~\bibnamefont {Rossignolo}}, \bibinfo {author}
  {\bibfnamefont {P.}~\bibnamefont {Rembold}}, \bibinfo {author} {\bibfnamefont
  {S.}~\bibnamefont {Montangero}}, \bibinfo {author} {\bibfnamefont
  {T.}~\bibnamefont {Calarco}}, \bibinfo {author} {\bibfnamefont
  {M.}~\bibnamefont {Endres}}, \bibinfo {author} {\bibfnamefont
  {M.}~\bibnamefont {Greiner}}, \bibinfo {author} {\bibfnamefont
  {V.}~\bibnamefont {Vuleti{\'c}}}, \ and\ \bibinfo {author} {\bibfnamefont
  {M.~D.}\ \bibnamefont {Lukin}},\ }\href {\doibase 10.1126/science.aax9743}
  {\bibfield  {journal} {\bibinfo  {journal} {Science}\ }\textbf {\bibinfo
  {volume} {365}},\ \bibinfo {pages} {570} (\bibinfo {year}
  {2019})}\BibitemShut {NoStop}%
\bibitem [{\citenamefont {Khaneja}\ \emph {et~al.}(2005)\citenamefont
  {Khaneja}, \citenamefont {Reiss}, \citenamefont {Kehlet}, \citenamefont
  {Schulte-Herbr\"uggen},\ and\ \citenamefont {Glaser}}]{KhanejaJMR05}%
  \BibitemOpen
  \bibfield  {author} {\bibinfo {author} {\bibfnamefont {N.}~\bibnamefont
  {Khaneja}}, \bibinfo {author} {\bibfnamefont {T.}~\bibnamefont {Reiss}},
  \bibinfo {author} {\bibfnamefont {C.}~\bibnamefont {Kehlet}}, \bibinfo
  {author} {\bibfnamefont {T.}~\bibnamefont {Schulte-Herbr\"uggen}}, \ and\
  \bibinfo {author} {\bibfnamefont {S.~J.}\ \bibnamefont {Glaser}},\ }\href
  {\doibase 10.1016/j.jmr.2004.11.004} {\bibfield  {journal} {\bibinfo
  {journal} {J. Magn. Reson.}\ }\textbf {\bibinfo {volume} {172}},\ \bibinfo
  {pages} {296 } (\bibinfo {year} {2005})}\BibitemShut {NoStop}%
\bibitem [{\citenamefont {Reich}\ \emph {et~al.}(2012)\citenamefont {Reich},
  \citenamefont {Ndong},\ and\ \citenamefont {Koch}}]{JCP.136.104103}%
  \BibitemOpen
  \bibfield  {author} {\bibinfo {author} {\bibfnamefont {D.~M.}\ \bibnamefont
  {Reich}}, \bibinfo {author} {\bibfnamefont {M.}~\bibnamefont {Ndong}}, \ and\
  \bibinfo {author} {\bibfnamefont {C.~P.}\ \bibnamefont {Koch}},\ }\href
  {\doibase http://dx.doi.org/10.1063/1.3691827} {\bibfield  {journal}
  {\bibinfo  {journal} {J. Chem. Phys.}\ }\textbf {\bibinfo {volume} {136}},\
  \bibinfo {eid} {104103} (\bibinfo {year} {2012})}\BibitemShut {NoStop}%
\bibitem [{\citenamefont {Goerz}\ \emph {et~al.}(2019)\citenamefont {Goerz},
  \citenamefont {Basilewitsch}, \citenamefont {Gago-Encinas}, \citenamefont
  {Krauss}, \citenamefont {Horn}, \citenamefont {Reich},\ and\ \citenamefont
  {Koch}}]{GoerzSciPost19}%
  \BibitemOpen
  \bibfield  {author} {\bibinfo {author} {\bibfnamefont {M.~H.}\ \bibnamefont
  {Goerz}}, \bibinfo {author} {\bibfnamefont {D.}~\bibnamefont {Basilewitsch}},
  \bibinfo {author} {\bibfnamefont {F.}~\bibnamefont {Gago-Encinas}}, \bibinfo
  {author} {\bibfnamefont {M.~G.}\ \bibnamefont {Krauss}}, \bibinfo {author}
  {\bibfnamefont {K.~P.}\ \bibnamefont {Horn}}, \bibinfo {author}
  {\bibfnamefont {D.~M.}\ \bibnamefont {Reich}}, \ and\ \bibinfo {author}
  {\bibfnamefont {C.~P.}\ \bibnamefont {Koch}},\ }\href {\doibase
  10.21468/SciPostPhys.7.6.080} {\bibfield  {journal} {\bibinfo  {journal}
  {SciPost Phys.}\ }\textbf {\bibinfo {volume} {7}},\ \bibinfo {pages} {80}
  (\bibinfo {year} {2019})}\BibitemShut {NoStop}%
\bibitem [{\citenamefont {Machnes}\ \emph {et~al.}(2018)\citenamefont
  {Machnes}, \citenamefont {Ass\'emat}, \citenamefont {Tannor},\ and\
  \citenamefont {Wilhelm}}]{MachnesPRL18}%
  \BibitemOpen
  \bibfield  {author} {\bibinfo {author} {\bibfnamefont {S.}~\bibnamefont
  {Machnes}}, \bibinfo {author} {\bibfnamefont {E.}~\bibnamefont {Ass\'emat}},
  \bibinfo {author} {\bibfnamefont {D.}~\bibnamefont {Tannor}}, \ and\ \bibinfo
  {author} {\bibfnamefont {F.~K.}\ \bibnamefont {Wilhelm}},\ }\href {\doibase
  10.1103/PhysRevLett.120.150401} {\bibfield  {journal} {\bibinfo  {journal}
  {Phys. Rev. Lett.}\ }\textbf {\bibinfo {volume} {120}},\ \bibinfo {pages}
  {150401} (\bibinfo {year} {2018})}\BibitemShut {NoStop}%
\bibitem [{\citenamefont {Doria}\ \emph {et~al.}(2011)\citenamefont {Doria},
  \citenamefont {Calarco},\ and\ \citenamefont {Montangero}}]{DoriaPRL11}%
  \BibitemOpen
  \bibfield  {author} {\bibinfo {author} {\bibfnamefont {P.}~\bibnamefont
  {Doria}}, \bibinfo {author} {\bibfnamefont {T.}~\bibnamefont {Calarco}}, \
  and\ \bibinfo {author} {\bibfnamefont {S.}~\bibnamefont {Montangero}},\
  }\href {\doibase 10.1103/PhysRevLett.106.190501} {\bibfield  {journal}
  {\bibinfo  {journal} {Phys. Rev. Lett.}\ }\textbf {\bibinfo {volume} {106}},\
  \bibinfo {pages} {190501} (\bibinfo {year} {2011})}\BibitemShut {NoStop}%
\bibitem [{\citenamefont {Caneva}\ \emph {et~al.}(2011)\citenamefont {Caneva},
  \citenamefont {Calarco},\ and\ \citenamefont {Montangero}}]{CanevaPRA11}%
  \BibitemOpen
  \bibfield  {author} {\bibinfo {author} {\bibfnamefont {T.}~\bibnamefont
  {Caneva}}, \bibinfo {author} {\bibfnamefont {T.}~\bibnamefont {Calarco}}, \
  and\ \bibinfo {author} {\bibfnamefont {S.}~\bibnamefont {Montangero}},\
  }\href {\doibase 10.1103/PhysRevA.84.022326} {\bibfield  {journal} {\bibinfo
  {journal} {Phys. Rev. A}\ }\textbf {\bibinfo {volume} {84}},\ \bibinfo
  {pages} {022326} (\bibinfo {year} {2011})}\BibitemShut {NoStop}%
\bibitem [{\citenamefont {Helstrom}(1976)}]{HELS67}%
  \BibitemOpen
  \bibfield  {author} {\bibinfo {author} {\bibfnamefont {C.~W.}\ \bibnamefont
  {Helstrom}},\ }\href@noop {} {\emph {\bibinfo {title} {Quantum Detection and
  Estimation Theory}}}\ (\bibinfo  {publisher} {Academic Press},\ \bibinfo
  {year} {1976})\BibitemShut {NoStop}%
\bibitem [{\citenamefont {Holevo}(1982)}]{HOLE82}%
  \BibitemOpen
  \bibfield  {author} {\bibinfo {author} {\bibfnamefont {A.~S.}\ \bibnamefont
  {Holevo}},\ }\href@noop {} {\emph {\bibinfo {title} {Probabilistic and
  Statistical Aspect of Quantum Theory}}}\ (\bibinfo  {publisher}
  {North-Holland},\ \bibinfo {year} {1982})\BibitemShut {NoStop}%
\bibitem [{\citenamefont {{Yuen}}\ \emph {et~al.}(1975)\citenamefont {{Yuen}},
  \citenamefont {{Kennedy}},\ and\ \citenamefont {{Lax}}}]{Yuen1975}%
  \BibitemOpen
  \bibfield  {author} {\bibinfo {author} {\bibfnamefont {H.}~\bibnamefont
  {{Yuen}}}, \bibinfo {author} {\bibfnamefont {R.}~\bibnamefont {{Kennedy}}}, \
  and\ \bibinfo {author} {\bibfnamefont {M.}~\bibnamefont {{Lax}}},\ }\href
  {\doibase 10.1109/TIT.1975.1055351} {\bibfield  {journal} {\bibinfo
  {journal} {IEEE Trans. Inf. Theory}\ }\textbf {\bibinfo {volume} {21}},\
  \bibinfo {pages} {125} (\bibinfo {year} {1975})}\BibitemShut {NoStop}%
\bibitem [{\citenamefont {Giovannetti}\ \emph {et~al.}(2011)\citenamefont
  {Giovannetti}, \citenamefont {Lloyd},\ and\ \citenamefont
  {Maccone}}]{giovannetti2011advances}%
  \BibitemOpen
  \bibfield  {author} {\bibinfo {author} {\bibfnamefont {V.}~\bibnamefont
  {Giovannetti}}, \bibinfo {author} {\bibfnamefont {S.}~\bibnamefont {Lloyd}},
  \ and\ \bibinfo {author} {\bibfnamefont {L.}~\bibnamefont {Maccone}},\
  }\href@noop {} {\bibfield  {journal} {\bibinfo  {journal} {Nat. Photonics}\
  }\textbf {\bibinfo {volume} {5}},\ \bibinfo {pages} {222} (\bibinfo {year}
  {2011})}\BibitemShut {NoStop}%
\bibitem [{\citenamefont {Giovannetti}\ \emph {et~al.}(2006)\citenamefont
  {Giovannetti}, \citenamefont {Lloyd},\ and\ \citenamefont
  {Maccone}}]{giovannetti2006quantum}%
  \BibitemOpen
  \bibfield  {author} {\bibinfo {author} {\bibfnamefont {V.}~\bibnamefont
  {Giovannetti}}, \bibinfo {author} {\bibfnamefont {S.}~\bibnamefont {Lloyd}},
  \ and\ \bibinfo {author} {\bibfnamefont {L.}~\bibnamefont {Maccone}},\
  }\href@noop {} {\bibfield  {journal} {\bibinfo  {journal} {Phys. Rev. Lett.}\
  }\textbf {\bibinfo {volume} {96}},\ \bibinfo {pages} {010401} (\bibinfo
  {year} {2006})}\BibitemShut {NoStop}%
\bibitem [{\citenamefont {Braunstein}\ \emph {et~al.}(1996)\citenamefont
  {Braunstein}, \citenamefont {Caves},\ and\ \citenamefont
  {Milburn}}]{braunstein1996generalized}%
  \BibitemOpen
  \bibfield  {author} {\bibinfo {author} {\bibfnamefont {S.~L.}\ \bibnamefont
  {Braunstein}}, \bibinfo {author} {\bibfnamefont {C.~M.}\ \bibnamefont
  {Caves}}, \ and\ \bibinfo {author} {\bibfnamefont {G.~J.}\ \bibnamefont
  {Milburn}},\ }\href@noop {} {\bibfield  {journal} {\bibinfo  {journal} {Ann.
  Phys.}\ }\textbf {\bibinfo {volume} {247}},\ \bibinfo {pages} {135} (\bibinfo
  {year} {1996})}\BibitemShut {NoStop}%
\bibitem [{\citenamefont {Fujiwara}\ and\ \citenamefont
  {Imai}(2008)}]{Fujiwara2008}%
  \BibitemOpen
  \bibfield  {author} {\bibinfo {author} {\bibfnamefont {A.}~\bibnamefont
  {Fujiwara}}\ and\ \bibinfo {author} {\bibfnamefont {H.}~\bibnamefont
  {Imai}},\ }\href {\doibase http://dx.doi.org/10.1088/1751-8113/41/25/255304}
  {\bibfield  {journal} {\bibinfo  {journal} {J. Phys. A}\ }\textbf {\bibinfo
  {volume} {41}},\ \bibinfo {pages} {255304} (\bibinfo {year}
  {2008})}\BibitemShut {NoStop}%
\bibitem [{\citenamefont {Escher}\ \emph {et~al.}(2011)\citenamefont {Escher},
  \citenamefont {de~Matos~Filho},\ and\ \citenamefont
  {Davidovich}}]{escher2012general}%
  \BibitemOpen
  \bibfield  {author} {\bibinfo {author} {\bibfnamefont {B.~M.}\ \bibnamefont
  {Escher}}, \bibinfo {author} {\bibfnamefont {R.~L.}\ \bibnamefont
  {de~Matos~Filho}}, \ and\ \bibinfo {author} {\bibfnamefont {L.}~\bibnamefont
  {Davidovich}},\ }\href {\doibase 10.1038/nphys1958} {\bibfield  {journal}
  {\bibinfo  {journal} {Nat. Phys.}\ }\textbf {\bibinfo {volume} {7}},\
  \bibinfo {pages} {406} (\bibinfo {year} {2011})}\BibitemShut {NoStop}%
\bibitem [{\citenamefont {Demkowicz-Dobrzanski}\ and\ \citenamefont
  {Maccone}(2014)}]{demkowicz2014using}%
  \BibitemOpen
  \bibfield  {author} {\bibinfo {author} {\bibfnamefont {R.}~\bibnamefont
  {Demkowicz-Dobrzanski}}\ and\ \bibinfo {author} {\bibfnamefont
  {L.}~\bibnamefont {Maccone}},\ }\href
  {http://link.aps.org/doi/10.1103/PhysRevLett.113.250801} {\bibfield
  {journal} {\bibinfo  {journal} {Phys. Rev. Lett.}\ }\textbf {\bibinfo
  {volume} {113}},\ \bibinfo {pages} {250801} (\bibinfo {year}
  {2014})}\BibitemShut {NoStop}%
\bibitem [{\citenamefont {Demkowicz-Dobrzanski}\ \emph
  {et~al.}(2012)\citenamefont {Demkowicz-Dobrzanski}, \citenamefont
  {Kolodynski},\ and\ \citenamefont {Guta}}]{demkowicz2012elusive}%
  \BibitemOpen
  \bibfield  {author} {\bibinfo {author} {\bibfnamefont {R.}~\bibnamefont
  {Demkowicz-Dobrzanski}}, \bibinfo {author} {\bibfnamefont {J.}~\bibnamefont
  {Kolodynski}}, \ and\ \bibinfo {author} {\bibfnamefont {M.}~\bibnamefont
  {Guta}},\ }\href {\doibase 10.1038/ncomms2067} {\bibfield  {journal}
  {\bibinfo  {journal} {Nat. Commun.}\ }\textbf {\bibinfo {volume} {3}},\
  \bibinfo {pages} {1063} (\bibinfo {year} {2012})}\BibitemShut {NoStop}%
\bibitem [{\citenamefont {Huelga}\ \emph {et~al.}(1997)\citenamefont {Huelga},
  \citenamefont {Macchiavello}, \citenamefont {Pellizzari}, \citenamefont
  {Ekert}, \citenamefont {Plenio},\ and\ \citenamefont
  {Cirac}}]{huelga1997improvement}%
  \BibitemOpen
  \bibfield  {author} {\bibinfo {author} {\bibfnamefont {S.~F.}\ \bibnamefont
  {Huelga}}, \bibinfo {author} {\bibfnamefont {C.}~\bibnamefont
  {Macchiavello}}, \bibinfo {author} {\bibfnamefont {T.}~\bibnamefont
  {Pellizzari}}, \bibinfo {author} {\bibfnamefont {A.~K.}\ \bibnamefont
  {Ekert}}, \bibinfo {author} {\bibfnamefont {M.~B.}\ \bibnamefont {Plenio}}, \
  and\ \bibinfo {author} {\bibfnamefont {J.~I.}\ \bibnamefont {Cirac}},\
  }\href@noop {} {\bibfield  {journal} {\bibinfo  {journal} {Phys. Rev. Lett.}\
  }\textbf {\bibinfo {volume} {79}},\ \bibinfo {pages} {3865} (\bibinfo {year}
  {1997})}\BibitemShut {NoStop}%
\bibitem [{\citenamefont {Chin}\ \emph {et~al.}(2012)\citenamefont {Chin},
  \citenamefont {Huelga},\ and\ \citenamefont {Plenio}}]{chin2012quantum}%
  \BibitemOpen
  \bibfield  {author} {\bibinfo {author} {\bibfnamefont {A.~W.}\ \bibnamefont
  {Chin}}, \bibinfo {author} {\bibfnamefont {S.~F.}\ \bibnamefont {Huelga}}, \
  and\ \bibinfo {author} {\bibfnamefont {M.~B.}\ \bibnamefont {Plenio}},\
  }\href@noop {} {\bibfield  {journal} {\bibinfo  {journal} {Phys. Rev. Lett.}\
  }\textbf {\bibinfo {volume} {109}},\ \bibinfo {pages} {233601} (\bibinfo
  {year} {2012})}\BibitemShut {NoStop}%
\bibitem [{\citenamefont {Berry}\ \emph {et~al.}(2015)\citenamefont {Berry},
  \citenamefont {Tsang}, \citenamefont {Hall},\ and\ \citenamefont
  {Wiseman}}]{Berry2015}%
  \BibitemOpen
  \bibfield  {author} {\bibinfo {author} {\bibfnamefont {D.~W.}\ \bibnamefont
  {Berry}}, \bibinfo {author} {\bibfnamefont {M.}~\bibnamefont {Tsang}},
  \bibinfo {author} {\bibfnamefont {M.~J.~W.}\ \bibnamefont {Hall}}, \ and\
  \bibinfo {author} {\bibfnamefont {H.~M.}\ \bibnamefont {Wiseman}},\ }\href
  {\doibase 10.1103/PhysRevX.5.031018} {\bibfield  {journal} {\bibinfo
  {journal} {Phys. Rev. X}\ }\textbf {\bibinfo {volume} {5}},\ \bibinfo {pages}
  {031018} (\bibinfo {year} {2015})}\BibitemShut {NoStop}%
\bibitem [{\citenamefont {Alipour}\ \emph {et~al.}(2014)\citenamefont
  {Alipour}, \citenamefont {Mehboudi},\ and\ \citenamefont
  {Rezakhani}}]{Alipour2014}%
  \BibitemOpen
  \bibfield  {author} {\bibinfo {author} {\bibfnamefont {S.}~\bibnamefont
  {Alipour}}, \bibinfo {author} {\bibfnamefont {M.}~\bibnamefont {Mehboudi}}, \
  and\ \bibinfo {author} {\bibfnamefont {A.~T.}\ \bibnamefont {Rezakhani}},\
  }\href {\doibase 10.1103/PhysRevLett.112.120405} {\bibfield  {journal}
  {\bibinfo  {journal} {Phys. Rev. Lett.}\ }\textbf {\bibinfo {volume} {112}},\
  \bibinfo {pages} {120405} (\bibinfo {year} {2014})}\BibitemShut {NoStop}%
\bibitem [{\citenamefont {Beau}\ and\ \citenamefont {del
  Campo}(2017)}]{Beau2017}%
  \BibitemOpen
  \bibfield  {author} {\bibinfo {author} {\bibfnamefont {M.}~\bibnamefont
  {Beau}}\ and\ \bibinfo {author} {\bibfnamefont {A.}~\bibnamefont {del
  Campo}},\ }\href {\doibase 10.1103/PhysRevLett.119.010403} {\bibfield
  {journal} {\bibinfo  {journal} {Phys. Rev. Lett.}\ }\textbf {\bibinfo
  {volume} {119}},\ \bibinfo {pages} {010403} (\bibinfo {year}
  {2017})}\BibitemShut {NoStop}%
\bibitem [{\citenamefont {Liu}\ \emph {et~al.}(2019)\citenamefont {Liu},
  \citenamefont {Yuan}, \citenamefont {Lu},\ and\ \citenamefont
  {Wang}}]{Liu2019}%
  \BibitemOpen
  \bibfield  {author} {\bibinfo {author} {\bibfnamefont {J.}~\bibnamefont
  {Liu}}, \bibinfo {author} {\bibfnamefont {H.}~\bibnamefont {Yuan}}, \bibinfo
  {author} {\bibfnamefont {X.-M.}\ \bibnamefont {Lu}}, \ and\ \bibinfo {author}
  {\bibfnamefont {X.}~\bibnamefont {Wang}},\ }\href {\doibase
  10.1088/1751-8121/ab5d4d} {\bibfield  {journal} {\bibinfo  {journal} {J.
  Phys. A}\ }\textbf {\bibinfo {volume} {53}},\ \bibinfo {pages} {023001}
  (\bibinfo {year} {2019})}\BibitemShut {NoStop}%
\bibitem [{\citenamefont {Tsang}\ \emph {et~al.}(2016)\citenamefont {Tsang},
  \citenamefont {Nair},\ and\ \citenamefont {Lu}}]{tsang2016quantum}%
  \BibitemOpen
  \bibfield  {author} {\bibinfo {author} {\bibfnamefont {M.}~\bibnamefont
  {Tsang}}, \bibinfo {author} {\bibfnamefont {R.}~\bibnamefont {Nair}}, \ and\
  \bibinfo {author} {\bibfnamefont {X.-M.}\ \bibnamefont {Lu}},\ }\href@noop {}
  {\bibfield  {journal} {\bibinfo  {journal} {Phys. Rev. X}\ }\textbf {\bibinfo
  {volume} {6}},\ \bibinfo {pages} {031033} (\bibinfo {year}
  {2016})}\BibitemShut {NoStop}%
\bibitem [{\citenamefont {{Ogawa}}\ and\ \citenamefont
  {{Nagaoka}}(2000)}]{Ogawa2000}%
  \BibitemOpen
  \bibfield  {author} {\bibinfo {author} {\bibfnamefont {T.}~\bibnamefont
  {{Ogawa}}}\ and\ \bibinfo {author} {\bibfnamefont {H.}~\bibnamefont
  {{Nagaoka}}},\ }\href {\doibase 10.1109/18.887855} {\bibfield  {journal}
  {\bibinfo  {journal} {IEEE Trans. Inf. Theory}\ }\textbf {\bibinfo {volume}
  {46}},\ \bibinfo {pages} {2428} (\bibinfo {year} {2000})}\BibitemShut
  {NoStop}%
\bibitem [{\citenamefont {{Ogawa}}\ and\ \citenamefont
  {{Hayashi}}(2004)}]{Ogawa2004}%
  \BibitemOpen
  \bibfield  {author} {\bibinfo {author} {\bibfnamefont {T.}~\bibnamefont
  {{Ogawa}}}\ and\ \bibinfo {author} {\bibfnamefont {M.}~\bibnamefont
  {{Hayashi}}},\ }\href {\doibase 10.1109/TIT.2004.828155} {\bibfield
  {journal} {\bibinfo  {journal} {IEEE Trans. Inf. Theory}\ }\textbf {\bibinfo
  {volume} {50}},\ \bibinfo {pages} {1368} (\bibinfo {year}
  {2004})}\BibitemShut {NoStop}%
\bibitem [{\citenamefont {Hayashi}(2002)}]{Hayashi2002}%
  \BibitemOpen
  \bibfield  {author} {\bibinfo {author} {\bibfnamefont {M.}~\bibnamefont
  {Hayashi}},\ }\href {\doibase 10.1088/0305-4470/35/50/307} {\bibfield
  {journal} {\bibinfo  {journal} {J. Phys. A}\ }\textbf {\bibinfo {volume}
  {35}},\ \bibinfo {pages} {10759} (\bibinfo {year} {2002})}\BibitemShut
  {NoStop}%
\bibitem [{\citenamefont {Audenaert}\ \emph {et~al.}(2007)\citenamefont
  {Audenaert}, \citenamefont {Calsamiglia}, \citenamefont {Mu\~noz Tapia},
  \citenamefont {Bagan}, \citenamefont {Masanes}, \citenamefont {Acin},\ and\
  \citenamefont {Verstraete}}]{QChernoff2007}%
  \BibitemOpen
  \bibfield  {author} {\bibinfo {author} {\bibfnamefont {K.~M.~R.}\
  \bibnamefont {Audenaert}}, \bibinfo {author} {\bibfnamefont {J.}~\bibnamefont
  {Calsamiglia}}, \bibinfo {author} {\bibfnamefont {R.}~\bibnamefont {Mu\~noz
  Tapia}}, \bibinfo {author} {\bibfnamefont {E.}~\bibnamefont {Bagan}},
  \bibinfo {author} {\bibfnamefont {L.}~\bibnamefont {Masanes}}, \bibinfo
  {author} {\bibfnamefont {A.}~\bibnamefont {Acin}}, \ and\ \bibinfo {author}
  {\bibfnamefont {F.}~\bibnamefont {Verstraete}},\ }\href {\doibase
  10.1103/PhysRevLett.98.160501} {\bibfield  {journal} {\bibinfo  {journal}
  {Phys. Rev. Lett.}\ }\textbf {\bibinfo {volume} {98}},\ \bibinfo {pages}
  {160501} (\bibinfo {year} {2007})}\BibitemShut {NoStop}%
\bibitem [{\citenamefont {Audenaert}\ \emph {et~al.}(2008)\citenamefont
  {Audenaert}, \citenamefont {Nussbaum}, \citenamefont {Szkola},\ and\
  \citenamefont {Verstraete}}]{Audenaert2008}%
  \BibitemOpen
  \bibfield  {author} {\bibinfo {author} {\bibfnamefont {K.~M.~R.}\
  \bibnamefont {Audenaert}}, \bibinfo {author} {\bibfnamefont {M.}~\bibnamefont
  {Nussbaum}}, \bibinfo {author} {\bibfnamefont {A.}~\bibnamefont {Szkola}}, \
  and\ \bibinfo {author} {\bibfnamefont {F.}~\bibnamefont {Verstraete}},\
  }\href {\doibase 10.1007/s00220-008-0417-5} {\bibfield  {journal} {\bibinfo
  {journal} {Commun. Math. Phys.}\ }\textbf {\bibinfo {volume} {279}},\
  \bibinfo {pages} {251} (\bibinfo {year} {2008})}\BibitemShut {NoStop}%
\bibitem [{\citenamefont {Hiai}\ and\ \citenamefont {Petz}(1991)}]{Hiai1991}%
  \BibitemOpen
  \bibfield  {author} {\bibinfo {author} {\bibfnamefont {F.}~\bibnamefont
  {Hiai}}\ and\ \bibinfo {author} {\bibfnamefont {D.}~\bibnamefont {Petz}},\
  }\href {\doibase 10.1007/BF02100287} {\bibfield  {journal} {\bibinfo
  {journal} {Commun. Math. Phys.}\ }\textbf {\bibinfo {volume} {143}},\
  \bibinfo {pages} {99} (\bibinfo {year} {1991})}\BibitemShut {NoStop}%
\bibitem [{\citenamefont {Hayashi}(2007)}]{Hayashi2007}%
  \BibitemOpen
  \bibfield  {author} {\bibinfo {author} {\bibfnamefont {M.}~\bibnamefont
  {Hayashi}},\ }\href {\doibase 10.1103/PhysRevA.76.062301} {\bibfield
  {journal} {\bibinfo  {journal} {Phys. Rev. A}\ }\textbf {\bibinfo {volume}
  {76}},\ \bibinfo {pages} {062301} (\bibinfo {year} {2007})}\BibitemShut
  {NoStop}%
\bibitem [{\citenamefont {Acin}(2001)}]{Acin2001}%
  \BibitemOpen
  \bibfield  {author} {\bibinfo {author} {\bibfnamefont {A.}~\bibnamefont
  {Acin}},\ }\href {\doibase 10.1103/PhysRevLett.87.177901} {\bibfield
  {journal} {\bibinfo  {journal} {Phys. Rev. Lett.}\ }\textbf {\bibinfo
  {volume} {87}},\ \bibinfo {pages} {177901} (\bibinfo {year}
  {2001})}\BibitemShut {NoStop}%
\bibitem [{\citenamefont {D'Ariano}\ \emph {et~al.}(2001)\citenamefont
  {D'Ariano}, \citenamefont {Lo~Presti},\ and\ \citenamefont
  {Paris}}]{Mauro2001}%
  \BibitemOpen
  \bibfield  {author} {\bibinfo {author} {\bibfnamefont {G.~M.}\ \bibnamefont
  {D'Ariano}}, \bibinfo {author} {\bibfnamefont {P.}~\bibnamefont {Lo~Presti}},
  \ and\ \bibinfo {author} {\bibfnamefont {M.~G.~A.}\ \bibnamefont {Paris}},\
  }\href {\doibase 10.1103/PhysRevLett.87.270404} {\bibfield  {journal}
  {\bibinfo  {journal} {Phys. Rev. Lett.}\ }\textbf {\bibinfo {volume} {87}},\
  \bibinfo {pages} {270404} (\bibinfo {year} {2001})}\BibitemShut {NoStop}%
\bibitem [{\citenamefont {Duan}\ \emph {et~al.}(2007)\citenamefont {Duan},
  \citenamefont {Feng},\ and\ \citenamefont {Ying}}]{Duan2007}%
  \BibitemOpen
  \bibfield  {author} {\bibinfo {author} {\bibfnamefont {R.}~\bibnamefont
  {Duan}}, \bibinfo {author} {\bibfnamefont {Y.}~\bibnamefont {Feng}}, \ and\
  \bibinfo {author} {\bibfnamefont {M.}~\bibnamefont {Ying}},\ }\href {\doibase
  10.1103/PhysRevLett.98.100503} {\bibfield  {journal} {\bibinfo  {journal}
  {Phys. Rev. Lett.}\ }\textbf {\bibinfo {volume} {98}},\ \bibinfo {pages}
  {100503} (\bibinfo {year} {2007})}\BibitemShut {NoStop}%
\bibitem [{\citenamefont {Lu}\ \emph {et~al.}(2012)\citenamefont {Lu},
  \citenamefont {Chen},\ and\ \citenamefont {Duan}}]{Cheng2012}%
  \BibitemOpen
  \bibfield  {author} {\bibinfo {author} {\bibfnamefont {C.}~\bibnamefont
  {Lu}}, \bibinfo {author} {\bibfnamefont {J.}~\bibnamefont {Chen}}, \ and\
  \bibinfo {author} {\bibfnamefont {R.}~\bibnamefont {Duan}},\ }\href
  {http://dl.acm.org/citation.cfm?id=2231036.2231045} {\bibfield  {journal}
  {\bibinfo  {journal} {Quantum Info. Comput.}\ }\textbf {\bibinfo {volume}
  {12}},\ \bibinfo {pages} {138} (\bibinfo {year} {2012})}\BibitemShut
  {NoStop}%
\bibitem [{\citenamefont {Chiribella}\ \emph {et~al.}(2008)\citenamefont
  {Chiribella}, \citenamefont {D'Ariano},\ and\ \citenamefont
  {Perinotti}}]{ChiribellaDP08}%
  \BibitemOpen
  \bibfield  {author} {\bibinfo {author} {\bibfnamefont {G.}~\bibnamefont
  {Chiribella}}, \bibinfo {author} {\bibfnamefont {G.~M.}\ \bibnamefont
  {D'Ariano}}, \ and\ \bibinfo {author} {\bibfnamefont {P.}~\bibnamefont
  {Perinotti}},\ }\href {\doibase 10.1103/PhysRevLett.101.180501} {\bibfield
  {journal} {\bibinfo  {journal} {Phys. Rev. Lett.}\ }\textbf {\bibinfo
  {volume} {101}},\ \bibinfo {pages} {180501} (\bibinfo {year}
  {2008})}\BibitemShut {NoStop}%
\bibitem [{\citenamefont {Duan}\ \emph {et~al.}(2009)\citenamefont {Duan},
  \citenamefont {Feng},\ and\ \citenamefont {Ying}}]{DuanFY09}%
  \BibitemOpen
  \bibfield  {author} {\bibinfo {author} {\bibfnamefont {R.}~\bibnamefont
  {Duan}}, \bibinfo {author} {\bibfnamefont {Y.}~\bibnamefont {Feng}}, \ and\
  \bibinfo {author} {\bibfnamefont {M.}~\bibnamefont {Ying}},\ }\href {\doibase
  10.1103/PhysRevLett.103.210501} {\bibfield  {journal} {\bibinfo  {journal}
  {Phys. Rev. Lett.}\ }\textbf {\bibinfo {volume} {103}},\ \bibinfo {pages}
  {210501} (\bibinfo {year} {2009})}\BibitemShut {NoStop}%
\bibitem [{\citenamefont {Harrow}\ \emph {et~al.}(2010)\citenamefont {Harrow},
  \citenamefont {Hassidim}, \citenamefont {Leung},\ and\ \citenamefont
  {Watrous}}]{Harrow2010}%
  \BibitemOpen
  \bibfield  {author} {\bibinfo {author} {\bibfnamefont {A.~W.}\ \bibnamefont
  {Harrow}}, \bibinfo {author} {\bibfnamefont {A.}~\bibnamefont {Hassidim}},
  \bibinfo {author} {\bibfnamefont {D.~W.}\ \bibnamefont {Leung}}, \ and\
  \bibinfo {author} {\bibfnamefont {J.}~\bibnamefont {Watrous}},\ }\href
  {\doibase 10.1103/PhysRevA.81.032339} {\bibfield  {journal} {\bibinfo
  {journal} {Phys. Rev. A}\ }\textbf {\bibinfo {volume} {81}},\ \bibinfo
  {pages} {032339} (\bibinfo {year} {2010})}\BibitemShut {NoStop}%
\bibitem [{\citenamefont {Duan}\ \emph {et~al.}(2008)\citenamefont {Duan},
  \citenamefont {Feng},\ and\ \citenamefont {Ying}}]{Duan2008}%
  \BibitemOpen
  \bibfield  {author} {\bibinfo {author} {\bibfnamefont {R.}~\bibnamefont
  {Duan}}, \bibinfo {author} {\bibfnamefont {Y.}~\bibnamefont {Feng}}, \ and\
  \bibinfo {author} {\bibfnamefont {M.}~\bibnamefont {Ying}},\ }\href {\doibase
  10.1103/PhysRevLett.100.020503} {\bibfield  {journal} {\bibinfo  {journal}
  {Phys. Rev. Lett.}\ }\textbf {\bibinfo {volume} {100}},\ \bibinfo {pages}
  {020503} (\bibinfo {year} {2008})}\BibitemShut {NoStop}%
\bibitem [{\citenamefont {Yuan}\ and\ \citenamefont
  {Fung}(2015)}]{yuan2015optimal}%
  \BibitemOpen
  \bibfield  {author} {\bibinfo {author} {\bibfnamefont {H.}~\bibnamefont
  {Yuan}}\ and\ \bibinfo {author} {\bibfnamefont {C.-H.~F.}\ \bibnamefont
  {Fung}},\ }\href@noop {} {\bibfield  {journal} {\bibinfo  {journal} {Phys.
  Rev. Lett.}\ }\textbf {\bibinfo {volume} {115}},\ \bibinfo {pages} {110401}
  (\bibinfo {year} {2015})}\BibitemShut {NoStop}%
\bibitem [{\citenamefont {Yuan}(2016)}]{yuan2016sequential}%
  \BibitemOpen
  \bibfield  {author} {\bibinfo {author} {\bibfnamefont {H.}~\bibnamefont
  {Yuan}},\ }\href@noop {} {\bibfield  {journal} {\bibinfo  {journal} {Phys.
  Rev. Lett.}\ }\textbf {\bibinfo {volume} {117}},\ \bibinfo {pages} {160801}
  (\bibinfo {year} {2016})}\BibitemShut {NoStop}%
\bibitem [{\citenamefont {Xu}\ \emph {et~al.}(2019)\citenamefont {Xu},
  \citenamefont {Li}, \citenamefont {Liu}, \citenamefont {Wang}, \citenamefont
  {Yuan},\ and\ \citenamefont {Wang}}]{Xu2019}%
  \BibitemOpen
  \bibfield  {author} {\bibinfo {author} {\bibfnamefont {H.}~\bibnamefont
  {Xu}}, \bibinfo {author} {\bibfnamefont {J.}~\bibnamefont {Li}}, \bibinfo
  {author} {\bibfnamefont {L.}~\bibnamefont {Liu}}, \bibinfo {author}
  {\bibfnamefont {Y.}~\bibnamefont {Wang}}, \bibinfo {author} {\bibfnamefont
  {H.}~\bibnamefont {Yuan}}, \ and\ \bibinfo {author} {\bibfnamefont
  {X.}~\bibnamefont {Wang}},\ }\href
  {https://doi.org/10.1038/s41534-019-0198-z} {\bibfield  {journal} {\bibinfo
  {journal} {npj Quantum Inf.}\ }\textbf {\bibinfo {volume} {5}} (\bibinfo
  {year} {2019})}\BibitemShut {NoStop}%
\bibitem [{\citenamefont {Liu}\ and\ \citenamefont
  {Yuan}(2017{\natexlab{a}})}]{Liu2017}%
  \BibitemOpen
  \bibfield  {author} {\bibinfo {author} {\bibfnamefont {J.}~\bibnamefont
  {Liu}}\ and\ \bibinfo {author} {\bibfnamefont {H.}~\bibnamefont {Yuan}},\
  }\href {\doibase 10.1103/PhysRevA.96.012117} {\bibfield  {journal} {\bibinfo
  {journal} {Phys. Rev. A}\ }\textbf {\bibinfo {volume} {96}},\ \bibinfo
  {pages} {012117} (\bibinfo {year} {2017}{\natexlab{a}})}\BibitemShut
  {NoStop}%
\bibitem [{\citenamefont {Liu}\ and\ \citenamefont
  {Yuan}(2017{\natexlab{b}})}]{Liu2017control}%
  \BibitemOpen
  \bibfield  {author} {\bibinfo {author} {\bibfnamefont {J.}~\bibnamefont
  {Liu}}\ and\ \bibinfo {author} {\bibfnamefont {H.}~\bibnamefont {Yuan}},\
  }\href@noop {} {\bibfield  {journal} {\bibinfo  {journal} {Phys. Rev. A}\
  }\textbf {\bibinfo {volume} {96}},\ \bibinfo {pages} {042114} (\bibinfo
  {year} {2017}{\natexlab{b}})}\BibitemShut {NoStop}%
\bibitem [{\citenamefont {Pang}\ and\ \citenamefont {Jordan}(2017)}]{Pang2017}%
  \BibitemOpen
  \bibfield  {author} {\bibinfo {author} {\bibfnamefont {S.}~\bibnamefont
  {Pang}}\ and\ \bibinfo {author} {\bibfnamefont {A.~N.}\ \bibnamefont
  {Jordan}},\ }\href {\doibase 10.1038/ncomms14695} {\bibfield  {journal}
  {\bibinfo  {journal} {Nat. Commun.}\ }\textbf {\bibinfo {volume} {8}},\
  \bibinfo {pages} {14695} (\bibinfo {year} {2017})}\BibitemShut {NoStop}%
\bibitem [{\citenamefont {Hou}\ \emph {et~al.}(2019)\citenamefont {Hou},
  \citenamefont {Wang}, \citenamefont {Tang}, \citenamefont {Yuan},
  \citenamefont {Xiang}, \citenamefont {Li},\ and\ \citenamefont
  {Guo}}]{Hou19control}%
  \BibitemOpen
  \bibfield  {author} {\bibinfo {author} {\bibfnamefont {Z.}~\bibnamefont
  {Hou}}, \bibinfo {author} {\bibfnamefont {R.-J.}\ \bibnamefont {Wang}},
  \bibinfo {author} {\bibfnamefont {J.-F.}\ \bibnamefont {Tang}}, \bibinfo
  {author} {\bibfnamefont {H.}~\bibnamefont {Yuan}}, \bibinfo {author}
  {\bibfnamefont {G.-Y.}\ \bibnamefont {Xiang}}, \bibinfo {author}
  {\bibfnamefont {C.-F.}\ \bibnamefont {Li}}, \ and\ \bibinfo {author}
  {\bibfnamefont {G.-C.}\ \bibnamefont {Guo}},\ }\href {\doibase
  10.1103/PhysRevLett.123.040501} {\bibfield  {journal} {\bibinfo  {journal}
  {Phys. Rev. Lett.}\ }\textbf {\bibinfo {volume} {123}},\ \bibinfo {pages}
  {040501} (\bibinfo {year} {2019})}\BibitemShut {NoStop}%
\bibitem [{\citenamefont {Naghiloo}\ \emph {et~al.}(2017)\citenamefont
  {Naghiloo}, \citenamefont {Jordan},\ and\ \citenamefont
  {Murch}}]{naghiloo2017achieving}%
  \BibitemOpen
  \bibfield  {author} {\bibinfo {author} {\bibfnamefont {M.}~\bibnamefont
  {Naghiloo}}, \bibinfo {author} {\bibfnamefont {A.}~\bibnamefont {Jordan}}, \
  and\ \bibinfo {author} {\bibfnamefont {K.}~\bibnamefont {Murch}},\
  }\href@noop {} {\bibfield  {journal} {\bibinfo  {journal} {Phys. Rev. Lett.}\
  }\textbf {\bibinfo {volume} {119}},\ \bibinfo {pages} {180801} (\bibinfo
  {year} {2017})}\BibitemShut {NoStop}%
\bibitem [{\citenamefont {Predko}\ \emph {et~al.}(2020)\citenamefont {Predko},
  \citenamefont {Albarelli},\ and\ \citenamefont {Serafini}}]{Predko2020}%
  \BibitemOpen
  \bibfield  {author} {\bibinfo {author} {\bibfnamefont {A.}~\bibnamefont
  {Predko}}, \bibinfo {author} {\bibfnamefont {F.}~\bibnamefont {Albarelli}}, \
  and\ \bibinfo {author} {\bibfnamefont {A.}~\bibnamefont {Serafini}},\
  }\href@noop {} {\bibfield  {journal} {\bibinfo  {journal} {arXiv:2001.03551}\
  } (\bibinfo {year} {2020})}\BibitemShut {NoStop}%
\bibitem [{\citenamefont {Mirkin}\ \emph {et~al.}(2019)\citenamefont {Mirkin},
  \citenamefont {Larocca},\ and\ \citenamefont {Wisniacki}}]{Mirkin2019}%
  \BibitemOpen
  \bibfield  {author} {\bibinfo {author} {\bibfnamefont {N.}~\bibnamefont
  {Mirkin}}, \bibinfo {author} {\bibfnamefont {M.}~\bibnamefont {Larocca}}, \
  and\ \bibinfo {author} {\bibfnamefont {D.}~\bibnamefont {Wisniacki}},\
  }\href@noop {} {\bibfield  {journal} {\bibinfo  {journal} {arXiv:1912.04675}\
  } (\bibinfo {year} {2019})}\BibitemShut {NoStop}%
\bibitem [{\citenamefont {Childs}\ \emph {et~al.}(2000)\citenamefont {Childs},
  \citenamefont {Preskill},\ and\ \citenamefont {Renes}}]{Childs2000}%
  \BibitemOpen
  \bibfield  {author} {\bibinfo {author} {\bibfnamefont {A.~M.}\ \bibnamefont
  {Childs}}, \bibinfo {author} {\bibfnamefont {J.}~\bibnamefont {Preskill}}, \
  and\ \bibinfo {author} {\bibfnamefont {J.}~\bibnamefont {Renes}},\ }\href
  {\doibase 10.1080/09500340008244034} {\bibfield  {journal} {\bibinfo
  {journal} {J. Mod. Opt.}\ }\textbf {\bibinfo {volume} {47}},\ \bibinfo
  {pages} {155} (\bibinfo {year} {2000})}\BibitemShut {NoStop}%
\bibitem [{\citenamefont {Chen}\ and\ \citenamefont {Yuan}(2019)}]{Chen2019}%
  \BibitemOpen
  \bibfield  {author} {\bibinfo {author} {\bibfnamefont {Y.}~\bibnamefont
  {Chen}}\ and\ \bibinfo {author} {\bibfnamefont {H.}~\bibnamefont {Yuan}},\
  }\href {\doibase 10.1103/PhysRevA.100.022336} {\bibfield  {journal} {\bibinfo
   {journal} {Phys. Rev. A}\ }\textbf {\bibinfo {volume} {100}},\ \bibinfo
  {pages} {022336} (\bibinfo {year} {2019})}\BibitemShut {NoStop}%
\bibitem [{\citenamefont {Van~Damme}\ \emph {et~al.}(2018)\citenamefont
  {Van~Damme}, \citenamefont {Ansel}, \citenamefont {Glaser},\ and\
  \citenamefont {Sugny}}]{PhysRevA.98.043421}%
  \BibitemOpen
  \bibfield  {author} {\bibinfo {author} {\bibfnamefont {L.}~\bibnamefont
  {Van~Damme}}, \bibinfo {author} {\bibfnamefont {Q.}~\bibnamefont {Ansel}},
  \bibinfo {author} {\bibfnamefont {S.~J.}\ \bibnamefont {Glaser}}, \ and\
  \bibinfo {author} {\bibfnamefont {D.}~\bibnamefont {Sugny}},\ }\href
  {\doibase 10.1103/PhysRevA.98.043421} {\bibfield  {journal} {\bibinfo
  {journal} {Phys. Rev. A}\ }\textbf {\bibinfo {volume} {98}},\ \bibinfo
  {pages} {043421} (\bibinfo {year} {2018})}\BibitemShut {NoStop}%
\bibitem [{\citenamefont {Caneva}\ \emph {et~al.}(2009)\citenamefont {Caneva},
  \citenamefont {Murphy}, \citenamefont {Calarco}, \citenamefont {Fazio},
  \citenamefont {Montangero}, \citenamefont {Giovannetti},\ and\ \citenamefont
  {Santoro}}]{CanevaPRL09}%
  \BibitemOpen
  \bibfield  {author} {\bibinfo {author} {\bibfnamefont {T.}~\bibnamefont
  {Caneva}}, \bibinfo {author} {\bibfnamefont {M.}~\bibnamefont {Murphy}},
  \bibinfo {author} {\bibfnamefont {T.}~\bibnamefont {Calarco}}, \bibinfo
  {author} {\bibfnamefont {R.}~\bibnamefont {Fazio}}, \bibinfo {author}
  {\bibfnamefont {S.}~\bibnamefont {Montangero}}, \bibinfo {author}
  {\bibfnamefont {V.}~\bibnamefont {Giovannetti}}, \ and\ \bibinfo {author}
  {\bibfnamefont {G.~E.}\ \bibnamefont {Santoro}},\ }\href {\doibase
  10.1103/PhysRevLett.103.240501} {\bibfield  {journal} {\bibinfo  {journal}
  {Phys. Rev. Lett.}\ }\textbf {\bibinfo {volume} {103}},\ \bibinfo {pages}
  {240501} (\bibinfo {year} {2009})}\BibitemShut {NoStop}%
\bibitem [{\citenamefont {Larrouy}\ \emph {et~al.}(2020)\citenamefont
  {Larrouy}, \citenamefont {Patsch}, \citenamefont {Richaud}, \citenamefont
  {Raimond}, \citenamefont {Brune}, \citenamefont {Koch},\ and\ \citenamefont
  {Gleyzes}}]{LarrouyPRX20}%
  \BibitemOpen
  \bibfield  {author} {\bibinfo {author} {\bibfnamefont {A.}~\bibnamefont
  {Larrouy}}, \bibinfo {author} {\bibfnamefont {S.}~\bibnamefont {Patsch}},
  \bibinfo {author} {\bibfnamefont {R.}~\bibnamefont {Richaud}}, \bibinfo
  {author} {\bibfnamefont {J.-M.}\ \bibnamefont {Raimond}}, \bibinfo {author}
  {\bibfnamefont {M.}~\bibnamefont {Brune}}, \bibinfo {author} {\bibfnamefont
  {C.~P.}\ \bibnamefont {Koch}}, \ and\ \bibinfo {author} {\bibfnamefont
  {S.}~\bibnamefont {Gleyzes}},\ }\href@noop {} {\bibfield  {journal} {\bibinfo
   {journal} {Phys. Rev. X}\ }\textbf {\bibinfo {volume} {10}},\ \bibinfo
  {pages} {in press} (\bibinfo {year} {2020})}\BibitemShut {NoStop}%
\bibitem [{\citenamefont {Schulte}\ \emph {et~al.}(2020)\citenamefont
  {Schulte}, \citenamefont {Mart{\'{i}}nez-Lahuerta}, \citenamefont
  {Scharnagl},\ and\ \citenamefont {Hammerer}}]{Schulte2020}%
  \BibitemOpen
  \bibfield  {author} {\bibinfo {author} {\bibfnamefont {M.}~\bibnamefont
  {Schulte}}, \bibinfo {author} {\bibfnamefont {V.~J.}\ \bibnamefont
  {Mart{\'{i}}nez-Lahuerta}}, \bibinfo {author} {\bibfnamefont {M.~S.}\
  \bibnamefont {Scharnagl}}, \ and\ \bibinfo {author} {\bibfnamefont
  {K.}~\bibnamefont {Hammerer}},\ }\href {\doibase 10.22331/q-2020-05-15-268}
  {\bibfield  {journal} {\bibinfo  {journal} {{Quantum}}\ }\textbf {\bibinfo
  {volume} {4}},\ \bibinfo {pages} {268} (\bibinfo {year} {2020})}\BibitemShut
  {NoStop}%
\bibitem [{\citenamefont {Martin}\ \emph {et~al.}(2019)\citenamefont {Martin},
  \citenamefont {Weigert},\ and\ \citenamefont {Giraud}}]{Martin2019}%
  \BibitemOpen
  \bibfield  {author} {\bibinfo {author} {\bibfnamefont {J.}~\bibnamefont
  {Martin}}, \bibinfo {author} {\bibfnamefont {S.}~\bibnamefont {Weigert}}, \
  and\ \bibinfo {author} {\bibfnamefont {O.}~\bibnamefont {Giraud}},\
  }\href@noop {} {\bibfield  {journal} {\bibinfo  {journal} {arXiv:1909.08355}\
  } (\bibinfo {year} {2019})}\BibitemShut {NoStop}%
\bibitem [{\citenamefont {Sadzak}\ \emph {et~al.}(2019)\citenamefont {Sadzak},
  \citenamefont {Carmele}, \citenamefont {Widmann}, \citenamefont {Nebel},
  \citenamefont {Knorr},\ and\ \citenamefont {Benson}}]{Sadzek2019}%
  \BibitemOpen
  \bibfield  {author} {\bibinfo {author} {\bibfnamefont {N.}~\bibnamefont
  {Sadzak}}, \bibinfo {author} {\bibfnamefont {A.}~\bibnamefont {Carmele}},
  \bibinfo {author} {\bibfnamefont {C.}~\bibnamefont {Widmann}}, \bibinfo
  {author} {\bibfnamefont {C.}~\bibnamefont {Nebel}}, \bibinfo {author}
  {\bibfnamefont {A.}~\bibnamefont {Knorr}}, \ and\ \bibinfo {author}
  {\bibfnamefont {O.}~\bibnamefont {Benson}},\ }\href@noop {} {\bibfield
  {journal} {\bibinfo  {journal} {arXiv:1912.09245}\ } (\bibinfo {year}
  {2019})}\BibitemShut {NoStop}%
\bibitem [{\citenamefont {Haine}\ and\ \citenamefont
  {Hope}(2020)}]{PhysRevLett.124.060402}%
  \BibitemOpen
  \bibfield  {author} {\bibinfo {author} {\bibfnamefont {S.~A.}\ \bibnamefont
  {Haine}}\ and\ \bibinfo {author} {\bibfnamefont {J.~J.}\ \bibnamefont
  {Hope}},\ }\href {\doibase 10.1103/PhysRevLett.124.060402} {\bibfield
  {journal} {\bibinfo  {journal} {Phys. Rev. Lett.}\ }\textbf {\bibinfo
  {volume} {124}},\ \bibinfo {pages} {060402} (\bibinfo {year}
  {2020})}\BibitemShut {NoStop}%
\bibitem [{\citenamefont {Breuer}\ and\ \citenamefont
  {Petruccione}(2002)}]{Breuer}%
  \BibitemOpen
  \bibfield  {author} {\bibinfo {author} {\bibfnamefont {H.-P.}\ \bibnamefont
  {Breuer}}\ and\ \bibinfo {author} {\bibfnamefont {F.}~\bibnamefont
  {Petruccione}},\ }\href@noop {} {\emph {\bibinfo {title} {The theory of open
  quantum systems}}},\ \bibinfo {edition} {1st}\ ed.\ (\bibinfo  {publisher}
  {Oxford University Press},\ \bibinfo {year} {2002})\BibitemShut {NoStop}%
\bibitem [{\citenamefont {Braunstein}\ and\ \citenamefont
  {Caves}(1994)}]{Brau94}%
  \BibitemOpen
  \bibfield  {author} {\bibinfo {author} {\bibfnamefont {S.~L.}\ \bibnamefont
  {Braunstein}}\ and\ \bibinfo {author} {\bibfnamefont {C.~M.}\ \bibnamefont
  {Caves}},\ }\href@noop {} {\bibfield  {journal} {\bibinfo  {journal} {Phys.
  Rev. Lett.}\ }\textbf {\bibinfo {volume} {72}},\ \bibinfo {pages} {3439}
  (\bibinfo {year} {1994})}\BibitemShut {NoStop}%
\bibitem [{\citenamefont {Jozsa}(1994)}]{Jozsa.JModOpt.41.2315}%
  \BibitemOpen
  \bibfield  {author} {\bibinfo {author} {\bibfnamefont {R.}~\bibnamefont
  {Jozsa}},\ }\href {\doibase 10.1080/09500349414552171} {\bibfield  {journal}
  {\bibinfo  {journal} {J. Mod. Opt.}\ }\textbf {\bibinfo {volume} {41}},\
  \bibinfo {pages} {2315} (\bibinfo {year} {1994})}\BibitemShut {NoStop}%
\bibitem [{\citenamefont {Nielsen}\ and\ \citenamefont
  {Chuang}(2011)}]{NielsenChuang}%
  \BibitemOpen
  \bibfield  {author} {\bibinfo {author} {\bibfnamefont {M.~A.}\ \bibnamefont
  {Nielsen}}\ and\ \bibinfo {author} {\bibfnamefont {I.~L.}\ \bibnamefont
  {Chuang}},\ }\href@noop {} {\emph {\bibinfo {title} {Quantum Computation and
  Quantum Information}}},\ \bibinfo {edition} {10th}\ ed.\ (\bibinfo
  {publisher} {Cambridge University Press},\ \bibinfo {year}
  {2011})\BibitemShut {NoStop}%
\bibitem [{\citenamefont {D'Alessandro}(2007)}]{DAlessandro}%
  \BibitemOpen
  \bibfield  {author} {\bibinfo {author} {\bibfnamefont {D.}~\bibnamefont
  {D'Alessandro}},\ }\href@noop {} {\emph {\bibinfo {title} {Introduction to
  Quantum Control and Dynamics}}},\ \bibinfo {edition} {1st}\ ed.\ (\bibinfo
  {publisher} {Chapman and Hall/CRC},\ \bibinfo {year} {2007})\BibitemShut
  {NoStop}%
\bibitem [{\citenamefont {Dodonov}\ \emph {et~al.}(2000)\citenamefont
  {Dodonov}, \citenamefont {Man'ko}, \citenamefont {Man'ko},\ and\
  \citenamefont {W\"unsche}}]{Dodonov.JModOPt.47.633}%
  \BibitemOpen
  \bibfield  {author} {\bibinfo {author} {\bibfnamefont {V.~V.}\ \bibnamefont
  {Dodonov}}, \bibinfo {author} {\bibfnamefont {O.~V.}\ \bibnamefont {Man'ko}},
  \bibinfo {author} {\bibfnamefont {V.~I.}\ \bibnamefont {Man'ko}}, \ and\
  \bibinfo {author} {\bibfnamefont {A.}~\bibnamefont {W\"unsche}},\ }\href
  {\doibase 10.1080/09500340008233385} {\bibfield  {journal} {\bibinfo
  {journal} {J. Mod. Opt.}\ }\textbf {\bibinfo {volume} {47}},\ \bibinfo
  {pages} {633} (\bibinfo {year} {2000})}\BibitemShut {NoStop}%
\bibitem [{\citenamefont {Xu}\ \emph {et~al.}(2004)\citenamefont {Xu},
  \citenamefont {Yan}, \citenamefont {Ohtsuki}, \citenamefont {Fujimura},\ and\
  \citenamefont {Rabitz}}]{Xu.JChemPhys.120.6600}%
  \BibitemOpen
  \bibfield  {author} {\bibinfo {author} {\bibfnamefont {R.}~\bibnamefont
  {Xu}}, \bibinfo {author} {\bibfnamefont {Y.}~\bibnamefont {Yan}}, \bibinfo
  {author} {\bibfnamefont {Y.}~\bibnamefont {Ohtsuki}}, \bibinfo {author}
  {\bibfnamefont {Y.}~\bibnamefont {Fujimura}}, \ and\ \bibinfo {author}
  {\bibfnamefont {H.}~\bibnamefont {Rabitz}},\ }\href {\doibase
  10.1063/1.1665486} {\bibfield  {journal} {\bibinfo  {journal} {J. Chem.
  Phys.}\ }\textbf {\bibinfo {volume} {120}},\ \bibinfo {pages} {6600}
  (\bibinfo {year} {2004})}\BibitemShut {NoStop}%
\bibitem [{\citenamefont {Basilewitsch}\ \emph {et~al.}(2019)\citenamefont
  {Basilewitsch}, \citenamefont {Koch},\ and\ \citenamefont
  {Reich}}]{BasilewitschAQT}%
  \BibitemOpen
  \bibfield  {author} {\bibinfo {author} {\bibfnamefont {D.}~\bibnamefont
  {Basilewitsch}}, \bibinfo {author} {\bibfnamefont {C.~P.}\ \bibnamefont
  {Koch}}, \ and\ \bibinfo {author} {\bibfnamefont {D.~M.}\ \bibnamefont
  {Reich}},\ }\href {\doibase 10.1002/qute.201800110} {\bibfield  {journal}
  {\bibinfo  {journal} {Adv. Quantum Technol.}\ }\textbf {\bibinfo {volume}
  {2}},\ \bibinfo {pages} {1800110} (\bibinfo {year} {2019})}\BibitemShut
  {NoStop}%
\bibitem [{\citenamefont {Konnov}\ and\ \citenamefont
  {Krotov}(1999)}]{AutomRemContr.60.1427}%
  \BibitemOpen
  \bibfield  {author} {\bibinfo {author} {\bibfnamefont {A.~I.}\ \bibnamefont
  {Konnov}}\ and\ \bibinfo {author} {\bibfnamefont {V.~F.}\ \bibnamefont
  {Krotov}},\ }\href@noop {} {\bibfield  {journal} {\bibinfo  {journal} {Autom.
  Rem. Contr.}\ }\textbf {\bibinfo {volume} {60}},\ \bibinfo {pages} {1427 }
  (\bibinfo {year} {1999})}\BibitemShut {NoStop}%
\bibitem [{\citenamefont {Haroche}\ and\ \citenamefont
  {Raimond}(2006)}]{HarocheRaimond}%
  \BibitemOpen
  \bibfield  {author} {\bibinfo {author} {\bibfnamefont {S.}~\bibnamefont
  {Haroche}}\ and\ \bibinfo {author} {\bibfnamefont {J.-M.}\ \bibnamefont
  {Raimond}},\ }\href@noop {} {\emph {\bibinfo {title} {Exploring the
  Quantum}}}\ (\bibinfo  {publisher} {Oxford University Press},\ \bibinfo
  {year} {2006})\BibitemShut {NoStop}%
\bibitem [{Note1()}]{Note1}%
  \BibitemOpen
  \bibinfo {note} {It takes roughly 100 iterations for Krotov's method to
  converge to these control fields. Employing e.g.\ the QDYN library~\cite
  {QDYN}, this takes less than a minute on a standard desktop
  computer.}\BibitemShut {Stop}%
\bibitem [{QDY()}]{QDYN}%
  \BibitemOpen
  \href@noop {} {}\bibinfo {note} {{QDYN library,
  \url{www.qdyn-library.net}}}\BibitemShut {NoStop}%
\end{thebibliography}

%

\end{document}